# Carbon Removal Capacity Estimation of Taiga Reforestation and Afforestation at the Western Boreal Edge Using Spatially Explicit Carbon Budget Modeling

Kevin Bradley Dsouza[1*], Enoch Ofosu[1], Richard Boudreault[1, 2, 3, 4], Juan Moreno-Cruz[5], and Yuri Leonenko[1, 6]

1 - Department of Earth and Environmental Sciences, University of Waterloo, Waterloo, Canada.

2 - Environmental Sustainability, Université de Sherbrooke, Sherbrooke, Canada.

3 - Department of Chemical Engineering and Civil, Geological and Mining Engineering, Polytechnique Montréal, Montréal, Canada. 4 - AWN Nanotech, Montréal, Canada.

5 - School of Environment, Enterprise and Development, University of Waterloo, Waterloo, Canada.

6 - Department of Geography and Environmental Management, University of Waterloo, Waterloo, Canada.

* - Correspondence: kevin.dsouza@uwaterloo.ca

## Abstract

Canada's northern boreal forest edge offers considerable potential for climate change mitigation through large-scale tree planting. Afforestation in these sparsely forested regions could assist the natural northward migration of forests while capitalizing on their carbon removal capacity. However, the sequestration potential is uncertain due to a lack of spatially explicit models. This study uses Monte Carlo estimates and a carbon budget model to quantify the carbon removal capacity of afforestation at the northwestern boreal edge from 2025 to 2100. We combined satellite inventory data and probabilistic fire regime representations to simulate total ecosystem carbon under scenarios considering fire return intervals, land classes, planting mortality, and climate variables. Our results indicate that afforesting ~6.4–32 million hectares could sequester ~3.88–19.4 Gigatonnes of $CO_2e$ over the next 75 years, with the Taiga Shield West ecozone showing the most potential. Even the conservative estimate (3.88 $GtCO_2e$) is over five times Canada's total annual greenhouse gas emissions, making it a substantial contribution toward the nation's 2050 net-zero goal. Further research is needed to refine these estimates, assess economic viability, and investigate impacts on regional processes like permafrost thaw and surface albedo.

## Introduction

Climate change is occurring at an accelerated pace in higher latitudes, with the Arctic warming about four times faster than the global average [1, 2]. This rapid warming poses significant challenges for northern ecosystems, but also creates opportunities for nature-based climate solutions (NbCS). Canada’s federal government aims to reduce national greenhouse gas (GHG) emissions by 40% below 2005 levels by 2030 and to achieve net zero emissions by 2050 [3]. Realizing these targets will require both industrial decarbonization and substantial contributions from NbCS to remove and store carbon. Canada’s boreal forests harbor vast carbon stocks, with managed boreal forests alone storing nearly 28 gigatonnes (Gt) of carbon, while unmanaged lands hold even larger quantities [4, 5].

However, the boreal forests are shifting northward, albeit slowly, driven by rising temperatures and retreating sea ice [6, 7]. Exactly how this shift will alter net carbon stocks remains largely uncertain because of the interplay between multiple ecosystem factors such as vegetation and albedo feedbacks, permafrost, and changing temperatures [6, 8]. Such a northward shift of the boreal has been observed by analysing inter-annual trends in annual maximum vegetation greenness using satellite observations, providing early indicators [7]; however, further analysis shows that this expansion of tree cover into the southern arctic is slow and is not sufficient to compensate for the rapid decline of tree cover in the southern boreal boundary due to wildfires and timber logging, indicating signs of biome contraction [9]. This duality, in which the boreal forest is both expanding northward and contracting in parts of its southern range, raises the question of whether targeted reforestation/afforestation efforts at the boreal–taiga interface could help stabilize and potentially expand this critical carbon sink.

Despite growing interest in boreal afforestation [10–12], major uncertainties persist. Boreal tree planting at northern latitudes ranks as a “high impact, high uncertainty” NbCS due to variables such as permafrost thaw, albedo feedbacks, site limitations, high wildfire frequency, and unpredictability in long-term carbon durability [10, 13]. It is crucial to address these uncertainties to better understand the role of northern boreal tree planting in meeting Canada's climate change mitigation targets (see section “Discussion”), however, an even more basic uncertainty from the perspective of climate change mitigation is the actual carbon removal capacity of reforestation/afforestation projects considering existing landscape details, changing climatic conditions, variations in possible scenarios, and spatially explicit modeling at appropriate ecozone scales, making it difficult to pinpoint which subregions and scenarios will provide the greatest removal capacity.

Historically, assessing this potential has been limited by methodological constraints. Many early carbon budget models were aspatial, relying on aggregated data that could not fully capture the fine-scale heterogeneity of ecosystems or the precise impact of localized disturbances [14-19]. Recent progress in the field has shifted towards spatially explicit frameworks capable of integrating high-resolution remote sensing data to model complex carbon dynamics across landscapes. The Generic Carbon Budget Model (GCBM) [11] represents a key advancement in this area. As an open-source, spatially explicit framework, the GCBM simulates all five IPCC-required carbon pools and is designed to process the large, geographic datasets essential for modern land-sector monitoring [11]. Its scientific foundation is rooted in the well-established CBM-CFS3 model [11], which is used for Canada's national forest carbon reporting (see section “Methods” for more details).

The GCBM’s ability to integrate satellite-based forest inventories and model disturbances at a granular level makes it uniquely suited to reduce the significant uncertainties in northern carbon removal estimates. For example, the model’s granularity allows for finer properties, such as specific land histories and probabilistic fire regimes, to be assigned to each grid cell. This detailed, cell-level approach provides the foundation for Monte Carlo simulations that generate more robust statistical estimates of carbon removal under various scenarios. To our knowledge, a spatially explicit model with this capability has not been previously applied to assess the carbon budget impacts of taiga reforestation and afforestation at the western boreal edge. Therefore, this study presents a novel application of the GCBM to provide a more robust, spatially detailed assessment of carbon dynamics in this critical, understudied region. Although national-scale analyses [14, 15] and few province-scale analyses [16-19] have evaluated mitigation potentials of boreal tree planting, few focus specifically on the boreal boundary, include factors such as region-specific fire regimes, conduct meaningful ablations, or consider spatially explicit inventory. We address this by considering the north-western boreal boundary belonging to the Taiga Plains (TP) and the Taiga Shield West (TSW) ecozones as our region of interest (see Box. 1a), and use the National Terrestrial Ecosystem Monitoring System (NTEMS) Satellite-Based Forest Inventory (SBFI) [20]. We use GCBM to conduct spatially explicit carbon budget modeling in a range of scenarios and provide monte carlo estimates of carbon removal capacity in these scenarios. For each scenario, which considers different land histories, planting mortality rates, and climatic variables, we run a suite of simulations to arrive at these estimates. We project forest growth using baseline and regional yield curves and incorporate probabilistic fire regimes to simulate natural disturbances. Our results provide insight into the role played by different relevant parameters in the eventual removal capacity, the outcomes resulting from the spread of scenarios, and how these can potentially be used to guide policy decisions [21].

## Results

In this study, we consider reforestation/afforestation in the free areas (see Methods) of the region of interest at the edge of the north-western boreal in 2025 and track changes in total ecosystem carbon (TEC) until 2100 (75 years). Hereafter, we refer collectively to reforestation and afforestation as afforestation, noting that historically forested lands undergoing afforestation represent reforestation. Subsequent sections examine carbon removal potential and investigate various experimental configurations to reveal interactions within the afforestation system. Unless stated otherwise, simulation parameters follow those detailed in the “Methods” section.

## The Fire Return Interval Exerts Strong Control on TEC

Fire is a critical disturbance factor in boreal forests, shaping forest age, composition, and diversity [22, 23]. The mean fire return interval (FRI), defined as the average years between consecutive fires at a given location, is the primary indicator used to characterize fire regimes [24-26]. Typically, mean FRI is derived by fitting a negative exponential distribution to the proportion of surviving forest [24, 27-29]. In contrast, this study samples FRI values directly from ecozone-specific Weibull distributions (Eq. 1 in Methods), parameterized using historical data [30, 31], which effectively capture observed FRIs across boreal ecozones [32-35]. For example, Landsat-based estimates suggest FRIs ranging from 150-500 years at the TP edge and 500-1500 years at the TSW edge of the northwestern boreal range [32]. We set our Weibull scale and shape parameters accordingly (see “Methods” and supplementary Fig. 1).

Total ecosystem carbon (TEC), the sum of carbon stored in aboveground biomass, belowground biomass, and dead organic matter, is reported as tonnes per hectare (see supplementary Fig. 2 for area distribution), considering only carbon accumulated during the experiment relative to initial conditions. Our results show that TEC at 2100 increases with longer FRIs but saturates near an FRI of 500 years (Box. 1b-d), consistent with our fire regime simulations (see "Methods"). Additionally, TEC at 2100 is significantly higher in historically forested land (FL) compared to historically non-forested land (NFL) following afforestation (Box. 1b). One explanation could be that previously forested areas inherently support forest growth better, whereas non-forested lands may require substantial interventions (e.g., soil amendments, nutrients) to achieve comparable carbon storage capacity (see supplementary section "Ablations with Land Classes, Soil Types, and Fire Regimes"). Afforestation in available areas within already forested land also substantially boosts TEC compared to a baseline scenario of allowing existing forests to grow without additional planting (Box. 1b). These patterns remain consistent across the explored FRIs (supplementary Fig. 3,4).

We further investigated the effects of poor site quality, limited afforestation success, and failed seeding strategies by imposing a generic 90% tree mortality rate. This simulates real-world scenarios, as >95% of Canadian afforestation occurs via planting rather than seeding due to high seeding failure rates [36]. Under these extreme mortality conditions, TEC at 2100 is only marginally better than baseline in historically forested land (Box. 1c). Conversely, in historically non-forested areas, introducing 90% mortality has minimal impact on TEC because survival rates are already low, making additional mortality less significant (Box. 1d). This relationship holds across the FRI ranges explored. To enhance robustness, future analyses could incorporate explicit fire-spread modeling (e.g., Burn-P3 [37]), accounting for ignition sources and wind or terrain effects. Additionally, integrating climate projections to examine how shifts in temperature and precipitation might alter FRI distributions could provide deeper insights into future TEC trajectories [38].

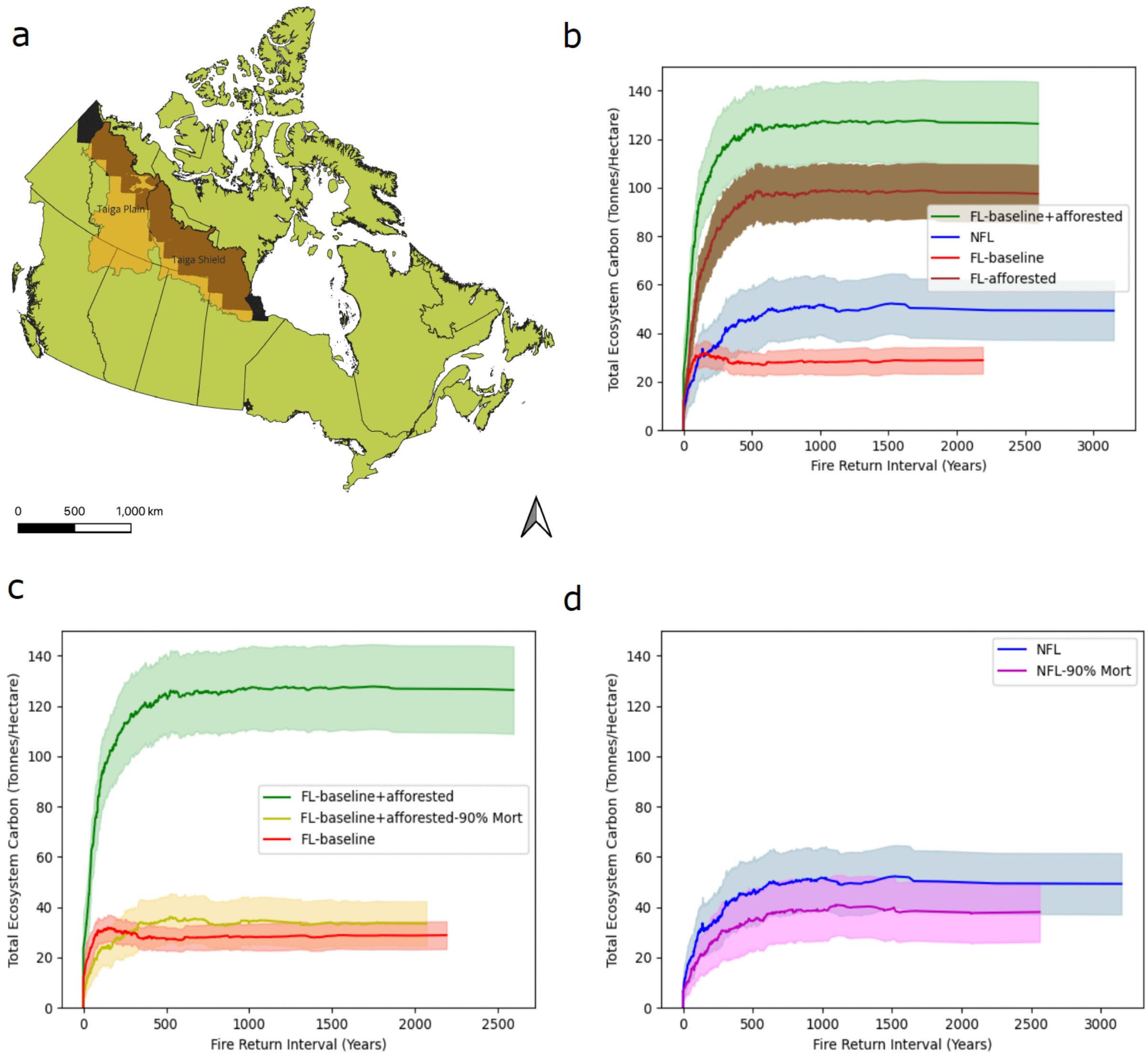

Box 1: **Region of interest and TEC at 2100 as a function of mean fire return interval (FRI). a)** Region of interest chosen from the NTEMS-SBFI. The strip was chosen from the NTEMS-SBFI [20] by considering the boreal forest edge in the north-west, spanning the TP and TSW ecozones, and the provinces of Northwest Territories (NT), Manitoba (MB), and Saskatchewan (SK). Yukon (YT), Nunavut (NU), and Taiga Taiga Cordillera (TC) were chosen, but were omitted for analysis because of issues with GCBM (see "Methods" section). **b)** TEC for afforestation on FL and NFL, and baseline and afforested configurations. **c)** TEC for afforestation on FL with and without generic mortality. 90% Mort refers to 90% mortality of afforested trees after 5 years. **d)** TEC for afforestation on NFL with and without generic mortality. FL - forested land, NFL - non-forested land. The baseline refers to existing forests without additional afforestation in free areas. The plotted values are averaged across a window of 100. The lines denote the mean and the spread shows the standard deviation computed across the 1000 independent simulation runs per scenario.

## The Effect of Generic Mortality on Ecosystem Carbon Growth

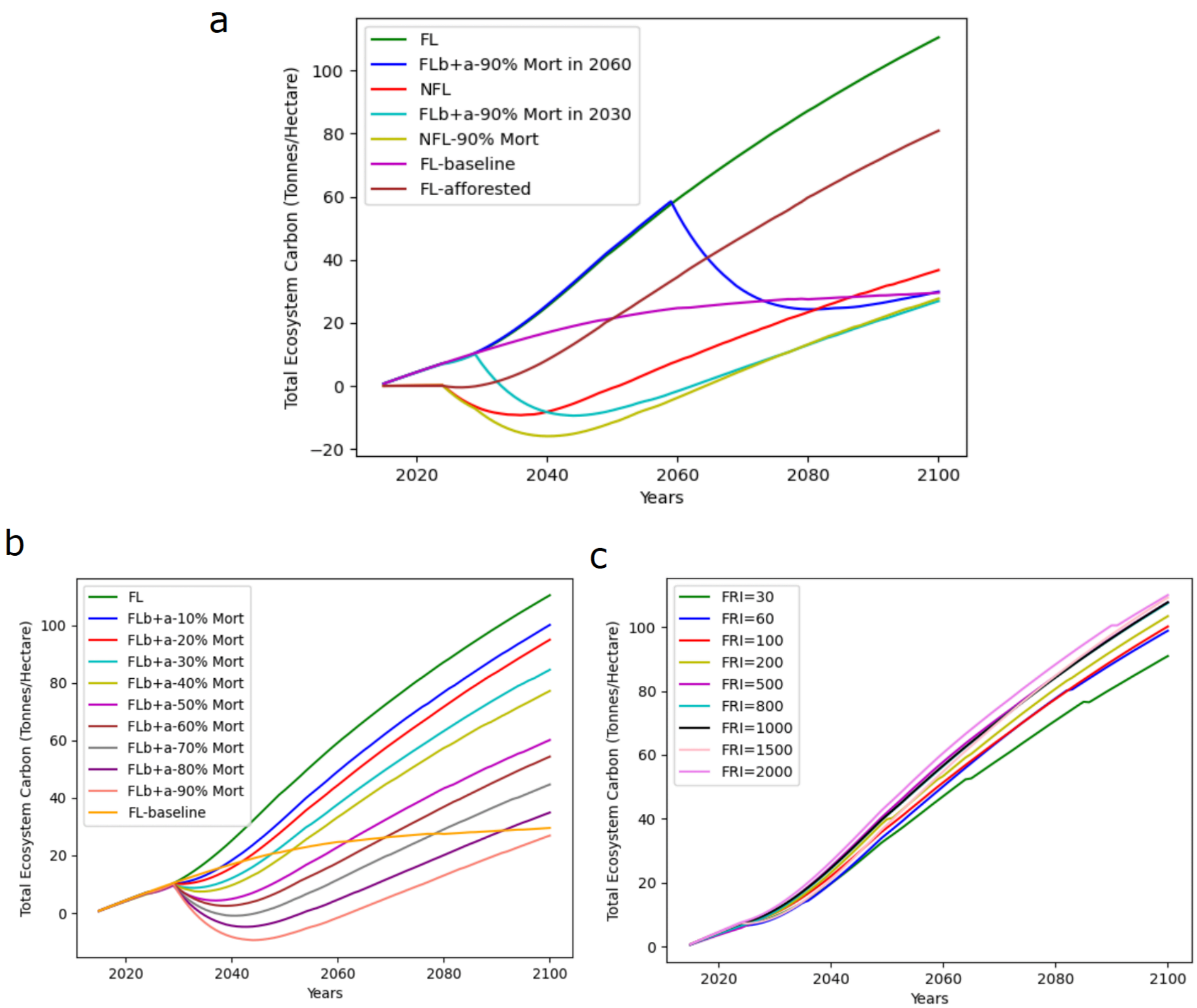

Box 2: **TEC over the years up to 2100. a)** FL - forested land, NFL - non-forested land, X% Mort - X% mortality of afforested trees after 5 years. The baseline refers to existing forests without additional afforestation in free areas. The plotted values are averaged across all other variables. The lines denote the means. The standard deviations can be found in supplementary Fig. 5. **b)** TEC with different percentages of mortality. Mortality acts as a control on the slope of TEC growth over the years. Standard deviation spread not shown to retain clarity. **c)** TEC with different mean fire return intervals (FRI) for afforestation on forested land (FL). The FRI affects the slope of TEC growth over the years to a lesser extent compared to mortality. Standard deviation spread not shown to retain clarity.

Examining TEC growth trajectories over time helps us understand how disturbances and historical land classes influence forest carbon accumulation up to 2100 [39, 40]. Initially, afforestation on NFL leads to reduced TEC, improving only later in the century (Box. 2a). This early decrease likely arises from poor growing conditions limiting biomass accumulation relative to disturbance-related losses [41]. Interestingly, applying high generic mortality (90%) results in similar TEC states by 2100 regardless of whether afforestation occurs on FL or NFL (Box. 2a), suggesting that at extreme mortality rates, the original land class becomes less influential [42]. Moreover, timing of mortality events (2030 vs. 2060) has little effect on TEC outcomes at equivalent mortality rates (Box. 2a). In contrast, the FL baseline scenario exhibits a plateau, whereas additional afforestation notably enhances TEC accumulation by 2100 (Box. 2a).

To better understand mortality's role in TEC growth, we varied generic mortality rates from 10% to 90%, observing a mostly linear effect on TEC growth, with minor deviations between 40% and 50% mortality possibly due to GCBM modeling artifacts (Box. 2b). Thus, mortality significantly regulates TEC growth slopes and can serve as an adjustable modeling parameter. Fire regimes, particularly fire return interval (FRI), also control TEC growth trajectories but have a smaller impact (Box. 2c; see section "The Fire Return Interval Exerts Strong Control on TEC"). Integrating species-specific mortality rates, distribution of mortality events spread out over the first few decades, and tying mortality to soil productivity or microclimate data, could be interesting extensions of this mortality analysis.

## Administrative-Ecozone Combinations and Climate Sensitivity

Our region of interest includes four key administrative-ecozone combinations: Northwest Territories (NT)-Taiga Shield West (TSW), NT-Taiga Plains (TP), Manitoba (MB)-TSW, and Saskatchewan (SK)-TSW. Separating the sequestration potential according to these combinations gives us an idea of how conducive each of these are for afforestation [4, 43]. We observe that NT-TSW has the highest sequestration potential as given by TEC at 2100 both for FL and NFL historical land classes (Box. 3a). While SK-TSW has the second highest TEC in FL, its potential in NFL is marginally better than MB-TSW, which is the lowest in NFL (Box. 3a). Climate change, altering mean annual temperature (MAT) and total annual precipitation (PCP), significantly impacts carbon sequestration. Direct MAT input to GCBM minimally changes TEC (see supplementary Fig. 6,7,8), indicating yield curves primarily account for climate effects in GCBM. Adjusting MAT and PCP in yield equations reveals increased sequestration only with rising temperatures (Box. 3b). An exception occurs in NT-TSW with reduced MAT and PCP (-20%) slightly outperforming averages (Box. 3b). Generally, lowest MAT (-40%) and highest PCP (+40%) scenarios reduce TEC the most (Box. 3b). Province-specific yield curves [44-47] and climate-driven disturbance regimes [38] could further refine these estimates.

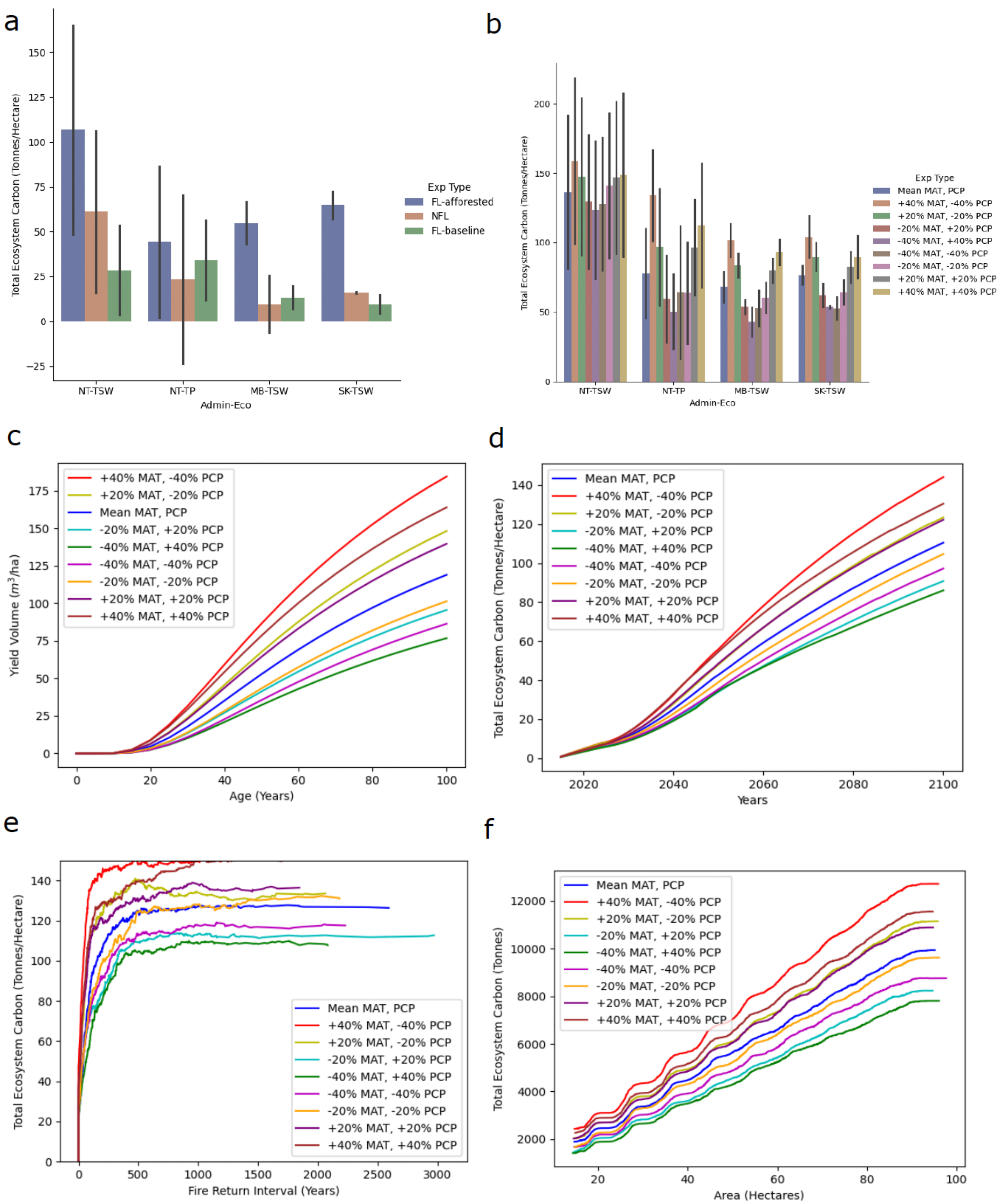

Box 3: **TEC at 2100 in different admin-eco combinations and with changes in MAT and PCP. a)** FL - forested land, NFL - non-forested land. The baseline refers to existing forests without additional afforestation in free areas. The plotted bars are averaged across all other variables. The bars denote the means and spread denotes the standard deviations computed across the 1000 independent simulation runs per scenario. **b)** TEC with different percentage changes in mean annual temperature (MAT) and total annual precipitation (PCP). The TEC accounts for existing plus afforested trees in FL. **c)** Yield volume as a function of age up to 100 years for species group 3 in the Taiga Plains eco-zone. **d)** TEC over the years up to 2100. The TEC accounts for existing plus afforested trees on forested land (FL). **e)** TEC at 2100 as a function of mean fire return interval (FRI). The plotted values are averaged across a window of 100. **f)** TEC at 2100 as a function of the afforestation area. A combination of existing simulations is sampled and combined to get higher afforestation area and the resulting TEC is added. Legend shows varying combination of increase or decrease in mean annual temperature (MAT) and total annual precipitation (PCP). Standard deviation spread not shown to retain clarity. The TEC accounts for existing plus afforested trees on forested land (FL).

Although the yield curves we use are parameterized by environmental variables, unbalanced sampling renders them a non-ideal candidate for faithfully modeling the impacts of climate change [48]. This is mainly due to the limited sampling in climate extremes and areas where climate is a limitation for tree growth, as well as the inability to handle non-linear effects [48]. However, climate sensitivity is a major limitation in most of existing Canadian growth and yield models, and obtaining reliable yield curves as a function of climate parameters is an ongoing effort [49-51]. Therefore, in this section, keeping in mind the limitations of our yield curves, we use them to stress test how the yield might behave with changing environmental conditions. Though this is not perfectly reliable, it allows us to speculate how the climate parameters might regulate growth in the future.

Altered yield curves (eq. 2 in section "Methods") for species group 3 in the TP eco-zone (for other species groups in the TP eco-zone see supplementary Fig. 9), illustrate that MAT strongly affects yield (higher MAT increases yield, lower MAT decreases yield), while PCP impacts yield modestly (Box. 3c). TEC trends closely follow these yield patterns, except for some specific MAT-PCP combinations (Box. 3d). Similar relationships appear between TEC, fire regimes, and afforestation area, emphasizing MAT's dominant role (Box. 3e,f). Future work should focus on refining climate-sensitive yield curves and modeling climate-influenced disturbances. Both the fire regime and the afforestation area scaling act as separate levers, acting in conjunction with the climate state (Box. 3e,f), however, the changing climate will certainly alter the fire regime and therefore restrict the space of possible scenarios. Altering the disturbance regimes with the changing climate and using proper climate-sensitive yield curves are natural extensions for future work.

## How Does Carbon Removal Capacity Scale with Afforestation Area

We run simulations on patches within gridded cells of approximately 7–11 hectares (supplementary Fig. 2) to maintain computational tractability. To explore how total ecosystem carbon (TEC) changes with larger afforestation areas, we combine afforestation areas and TEC from multiple independent experiments. Historical land class and mortality significantly influence TEC growth slopes relative to afforestation area (Box. 4a). Both FL and NFL with 90% mortality exhibit growth patterns similar to the FL-baseline scenario, whereas NFL without mortality shows marginally higher growth (Box. 4a). FL-afforested areas (part of FL-baseline+afforested) achieve the highest TEC growth rates (Box. 4a). In contrast, FRI has a lesser impact on TEC growth slopes on FL, except for the smallest FRI (FRI=30), which performs notably worse (Box. 4b). Other FRIs cluster closely, indicating similar growth patterns (Box. 4b). However, FRI significantly controls TEC growth on NFL (supplementary Fig. 10).

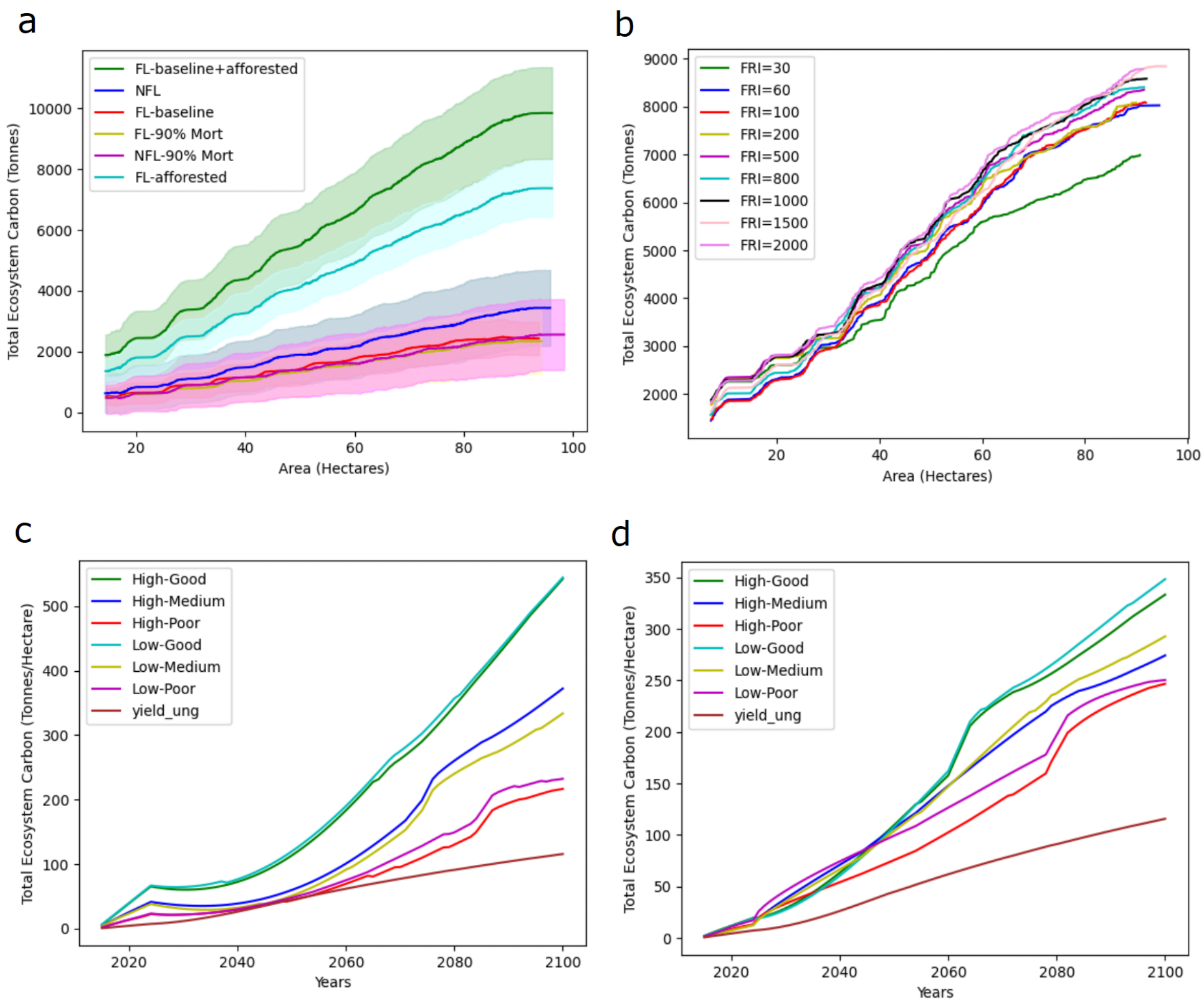

Box 4: **TEC at 2100 as a function of the afforestation area and over the years comparing yield curves by Timberworks Inc. and Ung et al. a)** FL - forested land, NFL - non-forested land, 90% Mort - 90% mortality of afforested trees after 5 years. The baseline refers to existing forests without additional afforestation in free areas. The lines denote the mean and the spread shows the standard deviation. **b)** TEC with different mean fire return intervals (FRI) for afforestation on forested land (FL). The TEC accounts for existing plus afforested trees in FL. A combination of existing simulations is sampled and combined to get higher afforestation area and the resulting TEC is added. Standard deviation spread not shown to retain clarity. **c)** TEC with high existing forest density. **d)** TEC with low existing forest density. The TEC accounts for existing plus afforested trees on FL. The yield curves from Ung et al. produce a lower bound for the TEC.

TEC increases roughly linearly with afforestation area when locations are sufficiently independent, suggesting greater TEC benefits from afforesting multiple smaller, independent patches rather than fewer correlated ones. Although our experiments combine patches randomly, deliberate selection minimizing parameter correlations may further enhance TEC growth slopes. Nonetheless, summing TEC from independently treated patches overlooks correlated risks, such as widespread wildfires [23], which could be addressed by explicitly modeling adjacency and connectivity effects [52, 53]. Additionally, our experiment does not capture non-linear, large-scale afforestation impacts on processes like competition, nutrient cycling, and hydrology [54], limitations stemming from GCBM constraints and patch combination methods. Resolving this is an interesting direction for future research.

## Practical Yield Curves and a Potential Lower Bound

In the previous sections, we used yield curves from Ung et al. [48] (eq. 2 in "Methods") to simulate forest growth. While these yield curves offer a useful starting point, being derived from sample plots across Canada, they may not fully represent our region due to local differences in site quality and climate. Additionally, these curves average across varying densities, limiting explicit analysis of forest density effects [48, 55-61]. To capture a realistic range of carbon sequestration estimates, we therefore consider alternative yield curves and their variability. We obtain yield curves from Timberworks Inc. in NT [62] and compare them against Ung et al. [48]. Timberworks Inc. used Alberta's Growth and Yield Projection System (GYPSY) to derive curves categorized by site productivity (good, medium, poor) and canopy density (dense or open) [62]. Supplementary Fig. 12 illustrates yield variations for NT across combinations of species, density, and site quality (Timberworks). Supplementary Figs. 9 and 13 provide corresponding data for TP and TSW using Ung et al.'s yield curves.

To explicitly analyze density effects, we simulated scenarios with high (Box. 4c) and low (Box. 4d) existing forest densities. We plot the TEC from the Ung et al. yield curve on both these sets of scenarios, and observe that it is a lower bound of the estimates (Box. 4c,d). Results indicate that site quality had a stronger influence on TEC than afforestation density, though higher initial density significantly increased overall TEC, likely due to faster carbon accumulation in mature, denser forests (Box. 4c,d). See supplementary section "Modulating Forest Density Using Mortality as a Surrogate" for additional analyses on forest density.

## Carbon Removal Capacity in the Taiga

We converted TEC into $CO_2$e removed (Tonnes), summarizing mean and standard deviation (computed across the 1000 independent simulation runs per scenario) for 2050, 2075, and 2100 (Box. 5a), assuming afforestation begins in 2025. Our analysis included afforestation on historically forested (FL) and non-forested lands (NFL), totaling approximately 6.4 M hectares, alongside existing forests covering ~8.6 M hectares. We see that the Ung et al. lower bound is much less compared to the average and best case carbon removal capacity obtained from yield curves from Timberworks (Box. 5a). By 2100, in the average case, afforesting 6.4 M hectares remove approximately 3.88 ± 0.98 Gt $CO_2$e, while existing forests remove 3.15 ± 0.92 Gt $CO_2$e. (Box. 5a). These estimates incorporate spatially explicit inventory data and fire regimes (see Methods), excluding additional mortality and climate change factors. Incorporating these factors requires using appropriate multiplicative adjustments. For instance, a 30% mortality within five years post-afforestation reduces carbon removal by approximately 20% (Box. 2b), whereas a warmer and drier climate scenario (+40% MAT, -40% PCP) increases carbon removal by around 35% (Box. 5b). The spatial distribution of this potential is illustrated in Box. 5d for Northwest Territories, the biggest region by area in our study.

Our estimate of 6.4 M hectares available for afforestation in northern regions is quite conservative. We assume that only ~15% of the total calculated free area in the study region (41.12 M hectares, see Box. 5c for a breakdown according to regions) is available, due to factors like logistical problems, biodiversity considerations, land ownership issues, and site amenability for afforestation. Moreover, we exclude territories like Yukon and Nunavut, the Taiga Cordillera ecozone, and the region immediately to the north of the Taiga. Given significant potential carbon removal in forest gaps and NFL regions of TP and TSW, evaluating the feasibility of taiga afforestation involves four key considerations: 1) future scenarios involving fire, mortality, disturbances, and climate changes that impact carbon sequestration [63]; 2) economic viability relative to alternative carbon removal strategies [64-66]; 3) ecological impacts including permafrost thaw, albedo feedbacks, energy fluxes, and ecosystem resilience [13]; and 4) Role of adaptive forest management strategies, including partial cutting, salvage logging, resilient species selection, and strategic fire management in maximizing long-term carbon storage. We plan to address these questions in detail in future research, however, Box. 5b shows an example of what the answer to the first question might look like.

We tabulate an illustrative set of ten scenarios (Box. 5b) to show how different parameters, fire regime (FRI), mortality rates, climate, and historical land class (forested [FL] vs. non-forested [NFL]), might affect total carbon ($CO_2e$) by 2100 across the 6.4 M ha considered (~15% of calculated free area, 3.2 M - FL, 3.2 M - NFL). We also compute best case estimates for when ~75% of the calculated free area may be available for afforestation. The "baseline" (Scenario 1) corresponds to the ~3.88 Gt $CO_2e$ figure derived from simulations using the Weibull-based FRI distributions for the Taiga Plains (TP) and Taiga Shield West (TSW), no early mortality, and no major climate departure. All other scenarios adjust key drivers and use approximate multipliers grounded in the relationships discussed in the Results section. Some key takeaways are: **a)** FL sequesters 1.8× more carbon per hectare than NFL in the baseline, so planting only FL or NFL yields about 1.38 Gt or 2.50 Gt, respectively, **b)** Moderate mortality (50%) reduces carbon by ~40%; extreme mortality (90%) cuts it to ~25% of baseline, **c)** Going from moderate, Weibull-based intervals to very short (~30 years) or very long (~500 years) can halve or boost carbon by 20%, **d)** +40% warming can increase overall yields by ~35%, whereas cooler scenarios (−20% MAT) cut carbon ~20%. However, the interaction between temperature and fire regimes and other disturbances are not taken into account here, as this is purely a yield curve based estimate, **e)** Multiplying individual factors can push net $CO_2e$ anywhere from <1 Gt in worst cases (frequent fires, extreme mortality) to over 6 Gt under ideal, low-disturbance, high-temperature scenarios.

We validated our GCBM projections against plot-level data from Canada's National Forest Inventory (NFI) [67]. The validation revealed a strong model performance in estimating carbon stocks and acceptable performance for predicting annual growth increments. As detailed in the "Validation Methodology" section in the Methods section, performance was assessed by comparing the mean NFI-derived values for each Age-Density-Species stratum against the corresponding mean of the outputs from GCBM. The overall $R^2$ and relative RMSE values across all strata are summarized in Box. 5e. The results align with expectations for forest carbon modelling. Our model demonstrated a good fit for Live Carbon Stocks ($R^2 = 0.65$), explaining 65% of the variance observed in the NFI data with a relatively low error (24%). Performance for TEC was lower ($R^2 = 0.57$), which is expected given the inclusion of the highly variable soil and deadwood pools. We consider the model's ability to predict the Live Carbon Increment ($R^2$ = 0.43) acceptable, as capturing the noisy, net result of annual growth and mortality is a significant challenge. Moreover, an $R^2$ value of 0.43 suggests that our modeling framework with GCBM correctly captures a good portion of the signal in this noisy process, successfully simulating the general trend and magnitude of forest growth dynamics in the different strata. Overall, these values fall within the ranges reported by other boreal forest carbon studies [68-70].

**a)**

| **Gt $CO_2$e Removed** | | **Ung et al. Lower Bound** | **Average** | **Best-Case** |
|---|---|---|---|---|
| **Mean - Afforestation** | 2050 | 0.14 | 0.47 | 0.76 |
| | 2075 | 0.71 | 2.32 | 3.72 |
| | 2100 | 1.19 | 3.88 | 6.2 |
| **Standard Deviation - Afforestation** | 2050 | 0.45 | 0.48 | 0.46 |
| | 2075 | 0.75 | 0.74 | 0.76 |
| | 2100 | 0.99 | 0.98 | 1.01 |
| **Mean - Existing Forests** | 2050 | 0.70 | 2.28 | 3.66 |
| | 2075 | 0.89 | 2.9 | 4.65 |
| | 2100 | 0.97 | 3.15 | 0.505 |
| **Standard Deviation - Existing Forests** | 2050 | 0.65 | 0.82 | 0.91 |
| | 2075 | 0.71 | 0.86 | 0.95 |
| | 2100 | 0.75 | 0.92 | 0.99 |

**b)**

| **Scenario** | **Key Parameters** | **Approx. Multipl** | **Resulting Gt $CO_2$e (~15% of** | **Resulting Gt $CO_2$e (~75% of available** | **Rationale/Multiplier Math** |
|---|---|---|---|---|---|

| | | ier | available area afforested) | area afforested) | |
|---|---|---|---|---|---|
| **Baseline** | - 6.4 M ha (3.2 M ha FL + 3.2 M ha NFL)<br>- Weibull-based FRIs (TP & TSW)<br>- No Mortality<br>- Current Climate | 1.00 | 3.88 | 19.4 | Starting point, combining both FL and NFL, modest disturbance regimes, and minimal seedling loss. |
| **Moderate Mortality** | - Same as Baseline except 50% Mortality (seed or establishment failures) | 0.60 | 2.33 | 11.65 | A ~40% reduction from Baseline: 3.88 × 0.60 ≈ 2.33 Gt. |
| **Extreme Mortality** | - Same as Baseline except 90% Mortality (severe early failures) | 0.25 | 0.97 | 4.85 | Only 25% of Baseline remains: 3.88 × 0.25 ≈ 0.97 Gt. |
| **FL Only** | - Afforest only 3.2 M ha of historically forested land | 0.64 | 2.47 | 12.35 | FL yields ~1.8× what NFL yields. NFL area = 3.2 M ha → total NFL carbon = 3.2x.<br>FL area = 3.2 M ha → total FL carbon = 3.2 × 1.8x = 5.76x.<br>Combined carbon = 3.2x + 5.76x = 8.96x = 3.88 Gt → x ≈ 0.43 Gt.<br>So, NFL portion ≈ 3.2x = 1.37 Gt, FL portion ≈ 5.76x = 2.47 Gt. 2.47 / 3.88 ≈ 0.64. |
| **NFL Only** | - Afforest only 3.2 M ha of historically non-forested land | 0.36 | 1.37 | 6.85 | NFL portion alone is 1.37 Gt; 1.37 / 3.88 ≈ 0.36. |
| **Short FRI** | - Same as Baseline except use a short FRI (≈30 years) | 0.60 | 2.33 | 11.65 | A ~40% reduction from Baseline: 3.88 × 0.60 ≈ 2.33 Gt. |
| **Long FRI** | - Same as Baseline except use a long FRI (≈500 years) | 1.20 | 4.66 | 23.3 | Reduced disturbance extends TEC accumulation by ~20%: 3.88 × 1.20 ≈ 4.66 Gt. |
| **Warmer Climate** | - Same as Baseline except ~+40% MAT (simulated in yield curves) | 1.35 | 5.24 | 26.2 | ~35% greater growth: 3.88 × 1.35 ≈ 5.24 Gt. |
| **Cooler Climate** | - Same as Baseline except ~−20% MAT (in yield curves) | 0.80 | 3.10 | 15.5 | ~20% decrease in growth: 3.88 × 0.80 ≈ 3.10 Gt. |
| **Long FRI + Warmer** | - FRI ~500 years (×1.20)<br>- +40% MAT (×1.35)<br>- No Mortality<br>- Full 6.4 M ha | 1.62 | 6.30 | 31.5 | Multiply both boosts: 3.88 × 1.20 × 1.35 ≈ 6.30 Gt. |

**c)**

| **Admin-zones** | **Eco-zones** | **Existing Forest Area (M hectares)** | **Afforestation Area: FL (M hectares)** | **Afforestation Area: NFL (M hectares)** |
|---|---|---|---|---|
| **Manitoba** | **Taiga Shield West** | 0.71 | 1.65 | 4.9 |
| **Northwest Territories** | **Taiga Plains** | 2.9 | 7.49 | 6.38 |
| **Northwest Territories** | **Taiga Shield West** | 4.86 | 11.36 | 8.96 |
| **Saskatchewan** | **Taiga Shield West** | 0.16 | 0.29 | 0.09 |
| | **Total Area** | 8.63 | 20.79 | 20.33 |

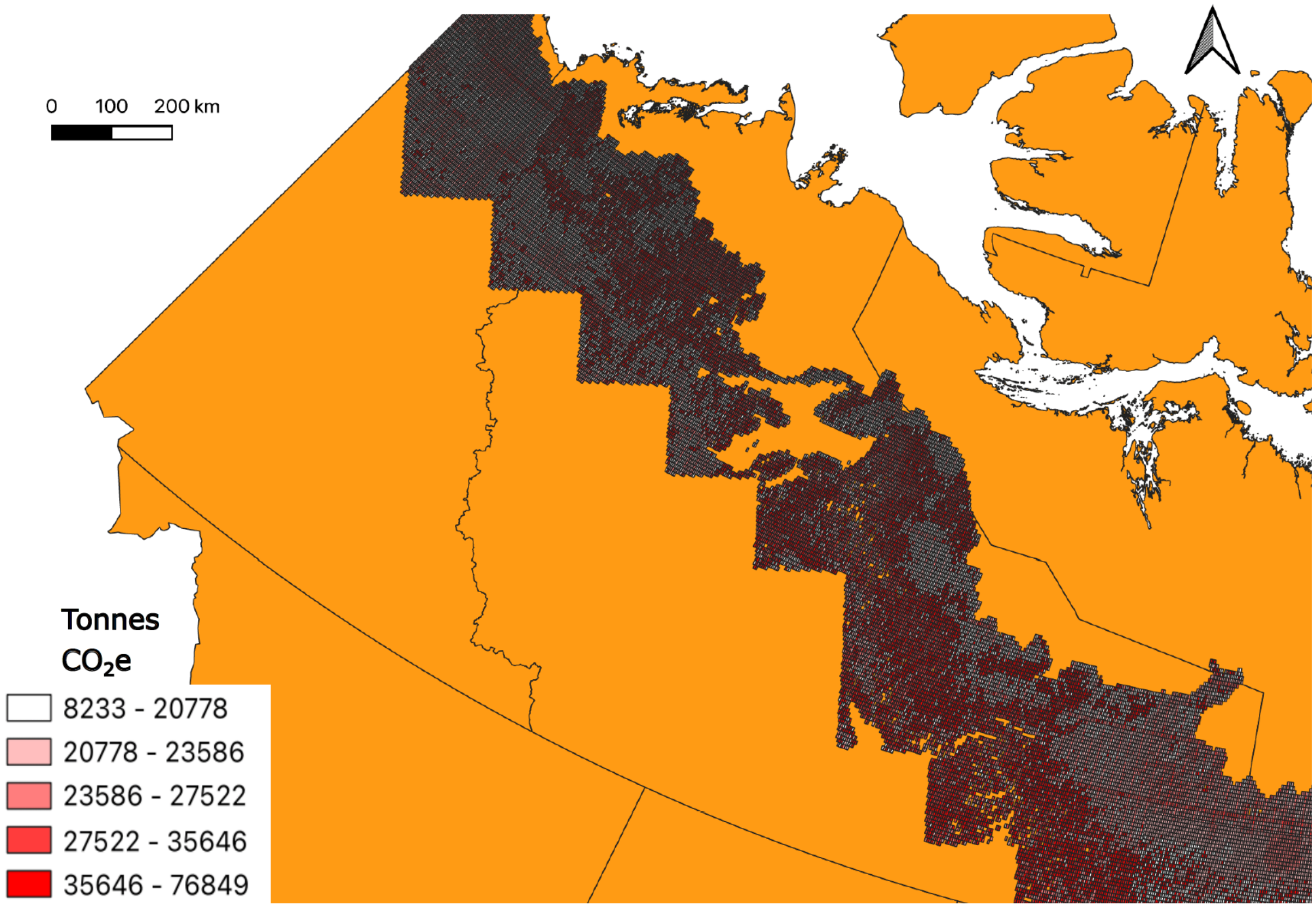

**d)**

**e)**

| Carbon Pool | $R^2$ (Coefficient of Determination) | RMSE |
|---|---|---|
| Live Carbon Stocks (AGB + BGB) | 0.65 | 24% |
| Total Ecosystem Carbon (TEC) Stocks | 0.57 | 33% |
| Live Carbon Increments | 0.43 | 39% |

Box 5: **Carbon removed via additional afforestation, existing forests, and different scenarios.** The total area afforested is ~6.4 M hectares. Mean and standard deviations shown for the Ung et al. lower bound, the average, and the best case scenarios. The average and best carbon removal capacity is calculated by using the different yield curves in NT for varying density and site quality by Timberworks Inc. [62]. **a)** Carbon removed in 2050, 2075, and 2100. **b)** Carbon removed in 2100 under ten illustrative scenarios considering fire regime (FRI), mortality rates, climate, and historical land classes. **c)** A breakdown of total areas considered in the study region according to different admin-eco combinations. We get a total of 41.12 M hectares of free area available for afforestation and 8.63M hectares of existing forests. **d)** Spatial map of average carbon removal capacity at 2100 in tons of $CO_2$e with additional afforestation, for Northwest Territories, the biggest region by area in our study. The afforested regions include both FL and NFL. The average carbon removal capacity is calculated by using the different yield curves in NT for varying density and site quality by Timberworks Inc. [62]. **e)** Overall model performance metrics for carbon stocks and increments. The comparison is between GCBM model outputs and mean values derived from NFI ground plots across all strata.

## Discussion

Our simulations demonstrate that fire return interval (FRI), historical land cover, climate parameters, and planting mortality collectively determine carbon removal from afforestation at the boreal–taiga interface. Longer FRIs greatly enhance total ecosystem carbon (TEC), as biomass accumulation has more time before fire resets succession. Conversely, short FRIs (<50 years), especially on historically non-forested land, limit carbon storage potential, underscoring the importance of effective fire management. Reforestation consistently surpasses afforestation on historically non-forested land due to better soil and microclimate conditions, although targeted soil amendments and species selection [71] may improve outcomes. High seedling mortality (around 90%) significantly undermines carbon gains, necessitating detailed site assessments, robust silviculture, and ongoing monitoring.

Yield curves showed sensitivity primarily to mean annual temperature (MAT), with precipitation (PCP) playing a secondary role. Moderate warming generally increases carbon removal, but extreme temperatures risk intensified fires or species stress, emphasizing the need for site-specific climate projections and calibrated yield models. Carbon accumulation scales linearly with afforestation area only if stands are relatively independent, suggesting a distributed approach to afforestation to mitigate risks from large disturbances. Incorporating density-specific yield curves, like those from GYPSY, TIPSY, and VDYP, might improve accuracy, addressing potential underestimation by default models [72]. For a deeper exploration of how historical land classes, soil types, and alternative fire-regime assignments influence total ecosystem carbon, see supplementary section “Ablations with Land Classes, Soil Types, and Fire Regimes”.

We don’t see organic soils play a dominant role in the big picture of TEC in our analyses. While GCBM includes soil organic carbon as a component of TEC, we don’t see noticeable differences in TEC between different soil types. A potential reason could be that the yield curves used to model tree growth do not directly account for interactions with soil type. Better integrating such complex interactions between growth models, soil characteristics, and climate is an active area of research for improving forest carbon modeling with GCBM. Another reason could be that the powerful and frequent effect of disturbances like wildfire, which resets ecosystem succession and releases large amounts of carbon, could overwhelm or mask the subtler, long-term influence of soil type on carbon accumulation. Finally, the simulations track carbon changes over a 75-year period, from 2025 to 2100. In slow-growing boreal ecosystems, the full impact of soil quality on biomass accumulation, and the subsequent transfer of carbon to dead organic matter and soil pools, may take longer than this timeframe to become statistically significant, especially when compared to the more immediate impacts of fire or planting mortality.

Although this study's quantitative findings are specific to the boreal taiga, its primary conclusions can be mapped onto the unique conditions of high-altitude regions [73, 74]. The critical influence of disturbance regimes like fire return intervals on carbon storage suggests that high-altitude disturbances, such as avalanches or pest outbreaks, will similarly govern sequestration success. Our finding that reforestation on historically forested land is far more effective than on non-forested sites implies that prioritizing high-altitude areas that recently lost tree cover would be more viable than attempting to plant in stable alpine meadows. The severe impact of high seedling mortality on carbon gains, the complex sensitivity of growth to climate variables, the risk-mitigation benefits of spatially diversifying plantations, and the necessity of weighing ecological trade-offs like altered albedo and hydrology are all directly transferable concepts that can inform more resilient and effective carbon sequestration strategies in high-altitude environments [73, 74].

The importance of this research extends beyond quantifying carbon sinks, positioning boreal afforestation within the broader portfolio of global climate solutions where it must be carefully weighed. As a Nature-based Climate Solution (NbCS), afforestation is a vital complement to, but not a substitute for, the primary goal of industrial decarbonization and reducing greenhouse gas emissions. When considered alongside technological options like direct air capture or other NbCS such as protecting existing forests, its implementation requires a clear-eyed assessment of its inherent challenges. These include significant financial and logistical costs in remote regions, the high uncertainty and risk of carbon reversal from disturbances like wildfire, and complex ecological trade-offs involving albedo feedbacks and regional hydrology. Therefore, studies like this one are crucial not to present afforestation as a standalone panacea, but to reduce its uncertainties, enabling policymakers to strategically deploy it where it is most cost-effective, resilient, and ecologically responsible as part of an integrated climate mitigation strategy.

Our findings provide policymakers with a scenario-based view of the potential carbon gains, and inherent uncertainties, associated with boreal afforestation. In particular, the results show the importance of aligning afforestation projects with strategic fire management (e.g., ensuring longer fire return intervals), selecting historically forested sites first (where growth is faster), and tailoring interventions to local climate projections. Quantifying how temperature and precipitation shifts, fire regimes, and early planting mortality impact overall carbon uptake enables governments and stakeholders to weigh costs and benefits more effectively against alternative climate mitigation strategies. Moreover, modeling different scenarios illuminates the range of possible outcomes, helping decision-makers identify “no-regrets” policy pathways, such as diversifying planting locations and optimizing species selection, that reduce the risk of large-scale carbon loss from fires, pests, or climatic extremes. Ultimately, this research offers a framework that can be integrated into broader land-use and climate policy discussions, guiding investments in northern afforestation where it can most sustainably and cost-effectively contribute to national and international carbon targets.

While afforestation supports natural boreal forest expansion, ecological and logistical challenges include limited infrastructure for seedling transport, sparse availability of nursery stock adapted to harsh boreal conditions, and higher costs of planting and maintenance in remote northern regions. Projected intensified wildfires could severely limit carbon retention, compromising large-scale planting efforts. By 2100, afforestation (~6.4 M hectares, conservative) could sequester around 3.88 Gt $CO_2$e, contributing significantly to Canada's climate goals, yet associated costs, wildfire uncertainty, and ecological trade-offs (albedo, permafrost) must be weighed. Forest canopies insulate permafrost [13], but lower albedo could enhance warming, emphasizing the need for regionally specific analyses [75-78] to identify optimal sites [79]. The management regime we use in our study is derived from the historical effects of management that are implicitly embedded within the empirical yield curves we use for projecting forest growth. However, adaptive forest management (e.g., partial cutting, salvage logging, resilient species selection) is crucial for mitigating leakage, baseline, and reversal risks, maximizing long-term carbon storage [80-85]. Future work could integrate such explicit management scenarios into our simulation framework. Future studies could also integrate province-specific yield curves, insect disturbances, and diverse management strategies (partial cutting, strategic harvesting). Explicit soil data from NSDB [86] and better modeling of cryosolic soils at the Arctic edge, alongside refined fire regime models (e.g., Burn-P3 [37]), will improve predictive accuracy for carbon sequestration (see supplementary section "Future Research").

While this study identifies significant afforestation potential by targeting "free areas," the methodology classifies ecologically active systems such as wetlands, shrublands, and grasslands as land available for planting. These non-forested ecosystems possess their own intrinsic biodiversity and provide critical ecological functions that would be lost upon conversion to forest [87]. For instance, the sedge meadows and floodplains identified as potential free areas for afforestation are crucial year-round grazing habitats for wild herds of Wood Bison. These same open wetlands provide essential breeding pools for amphibians like the Wood Frog and are nesting grounds for shorebirds such as the Lesser Yellowlegs. The open, lichen-rich peatlands and shrub barrens targeted for planting are critical summer foraging and calving grounds for Boreal Woodland Caribou. The unique mix of stunted trees and open tundra in this region is also the exclusive breeding habitat for endemic species like the Harris's Sparrow. Afforestation would directly compromise the survival of these species specifically adapted to open landscapes. Therefore, there is a conflict between maximizing carbon sequestration and conserving existing habitat, a trade-off that would practically limit the total area suitable for afforestation.

Our spatially explicit modeling of boreal–taiga afforestation at the north-western edge of Canada's boreal forest indicates that strategic planting could substantially increase total ecosystem carbon (TEC) by 2100. Longer fire return intervals, historically forested land cover, and low early seedling mortality all drive higher carbon gains. When scaled up, these findings emphasize that large swaths of Canada's northern boreal, especially areas with prior forest history, could play a more prominent role in climate mitigation portfolios. However, afforestation in these regions also faces risks from the very dynamics that define boreal ecosystems, notably wildfire and harsh winter conditions. For boreal planting initiatives to succeed long-term, careful site selection, adapted silvicultural practices, monitoring, and adaptive management will be paramount.

## Methods

The following sections detail the simulation softwares, data, pre-processing methods, and assumptions used for our spatially explicit carbon budget modeling.

## Generic Carbon Budget Model (GCBM)

The GCBM is a flexible, open-source framework for modeling forest carbon dynamics at the stand and landscape levels [11]. It generates a time-series output of spatially explicit and tabular indicators of forest carbon stocks and fluxes. Moreover, it adheres to the carbon estimation guidelines set by the Intergovernmental Panel on Climate Change (IPCC), simulating the dynamics of forest carbon stocks, including, aboveground biomass, belowground biomass, litter, dead wood, and soil organic carbon. The GCBM is meant to succeed the Canadian Forest Service's (CFS) CBM-CFS3 model in Canada's National Forest Carbon Monitoring, Accounting and Reporting System (NFCMARS) [88], which is responsible for tracking Canada's managed forest carbon balance for international reporting purposes. The GCBM is rooted in the same scientific foundation as the CBM-CFS3, but with the added benefit of spatial capabilities and enhanced data from national ecological parameter databases. While the CBM-CFS3 is an aspatial model, the GCBM requires spatial inputs, including forest inventory data and disturbance area information, to operate effectively. Notably, the GCBM is designed to integrate seamlessly with the Full Lands Integration Tool [89] software platform, developed by moja global. Recent studies have conducted accurate parameterization, robust uncertainty assessments, and sensitivity analyses, for CBM-CFS3, and by extension for GCBM [90-92]. Therefore, we don’t study the inherent parameter uncertainty in GCBM in detail here.

## Region of Interest

The north-western edge of the boreal forests is chosen as the region of interest from the National Terrestrial Ecosystem Monitoring System (NTEMS) Satellite-Based Forest Inventory (SBFI) [20] (Box. 1a). The NTEMS-SBFI data is available as rectangular layers, each consisting of multiple polygons. The grid layers along the edge of the forests are selected to be included in the inventory. The region of interest includes the eco-zones Taiga Plains (TP), Taiga Shield West (TSW), and Taiga Cordillera (TC), and the admin-zones Northwest Territories (NT), Yukon (YT), Manitoba (MB), Saskatchewan (SK), and Nunavut (NU).

## Compiling the NTEMS-SBFI data in the region of interest

The region of interest is gridded to 0.06 x 0.06 degrees (latitude, longitude) resolution and the polygons in the NTEMS-SBFI are assigned to these grid cells if they lie inside the cell or overlap with the cell. The data belonging to the polygons in each grid cell is converted to aggregate data for the grid cell. For instance, for a given cell, the tree species percentages are averaged across all the polygons in the cell and assigned to the cell. Similarly, forest age, fire fractions, percentage free area, and percentage forested area are averaged across all the polygons belonging to the cell. The historical fire year, admin-zone and eco-zone are assigned according to the most common occurrence among all the polygons in the cell. The mean average temperature (MAT) and total annual precipitation (PCP) are downloaded from ClimateDataCA [93], gridded to the same resolution as the grid cells (0.06 x 0.06 degrees), and assigned to them. Next, the gridded cells are retained if they belong to TP or TSW eco-zones. Currently, the TC eco-zone is not considered in the inventory because GCBM displays errors when running simulations in this eco-zone. The TC eco-zone will be considered in future experiments once these errors are resolved.

The resulting grid cells are further filtered depending on whether they have >35% free area. The free area is calculated by summing the area comprising exposed regions, bryoids, shrubs, herbs, and wetland, and excludes areas containing water, snow, rock, and forest. This is done to identify gaps along the boreal edge where afforestation can potentially be carried out. Therefore, the resulting estimates could be considered a lower bound because grid cells with <35% area are ignored and some snow covered regions which could be afforested are ignored. On the other hand, many exposed regions and regions with bryoids, shrubs, herbs, and wetland may not be amenable to afforestation. The conservative estimates balance these situations. Each grid cell in the retained inventory consists of the following relevant compiled data: a) percentages of different species on forested land, b) fraction of the region affected by fire, c) the year in which the historical fire disturbance occurred, d) admin-zone and eco-zone, e) mean forest age for forested region, f) percentage free and forested area, and d) MAT and PCP. For more details on the compiled forest inventory data in the region of interest, see section "Forest Inventory of the Region of Interest" in the "Methods" section.

## Creating scenarios

For an experiment with a chosen configuration, 1000 independent simulations per scenario configuration are created. We chose *n=1000* as we see that our variable ranges are sufficiently spanned at this *n* and it keeps simulations computationally tractable. If the experiment is decided to be run on historically forested land (FL), then the grid cells are filtered using the condition that the mean forest age in the cell is not zero and the percentage free area is <70%. If the experiment is decided to be run on historically non-forested land (NFL), cropland (CL), or grassland (GL), then the grid cells are filtered using the condition that the mean forest age in the cell is zero and the percentage free area is >=70%. Next, each scenario is created by doing the following:

**1)** A random cell is sampled from the grid, dividing the cell further into a 0.002 x 0.002 degrees grid, and sampling a sub-cell randomly. The sub-cell region is assigned as the inventory region and the historical land class for the inventory region is assigned according to the experiment type (either FL or NFL). **2)** If the historical land class is "FL", forest age is assigned to the inventory region by noting the mean forest age for the cell and sampling from a normal distribution with mean=mean forest age and standard deviation=0.1. If the historical land class is "NFL", forest age is assigned as 0. **3)** Eco-zone of the cell and the dominant soil type mapping for the eco-zone (Box. 6a) are noted. A dominant soil probability (*dom_soil_prob*) (see Box. 6b) is sampled randomly and the dominant soil type is assigned according to *dom_soil_prob* and “Average” soil type is assigned according to 1-dom_soil_prob. Here it would be ideal to include the full spatially explicit soil types from Soil Landscapes of Canada (SLC v3.2) [86], but this data is not used currently. **4)** Historic species are assigned to the inventory region by a weighted sampling according to species percentages in the cell. If the cell has no trees, a historical species of “Nope” is assigned. **5)** An afforestation region is chosen based on the percentage of free area in the cell. This region is assigned starting from the top portion of the inventory region, with lat range=percentage of free area*lat range of inventory region. The exact area with trees in the cell, combining both existing and afforested trees, is decided by GCBM and spans a distribution (see supplementary Fig. 2). **6)** The species to be afforested is chosen based on historical land class. If “FL”, the species is chosen via weighted sampling according to species percentages in the cell. If “NFL”, it is chosen via weighted sampling according to species percentages representative of the eco-zone of the cell (see Box. 6c). **7)** The admin-zone, eco-zone, and MAT of the cell are assigned to the inventory and afforestation regions. **8)** The simulation start and end years are set to 2015 and 2100 respectively. The afforestation year is set to 2025. If the experiment configuration involves generic mortality, generic mortality percentage and year of disturbance are set. Generic mortality is considered a proxy for planting strategy (perhaps also site quality), where seeding risks high mortality and transplanting results in low mortality. A similar relationship can be drawn for site quality. **9)** Fire events are assigned by sampling fire return intervals (FRIs) from a weibull distribution (Eq. 1). The weibull distribution parameter *scale* is sampled according to eco-zone specific mapping (see Box. 6d) based on the eco-zone of the cell. The weibull *shape* parameter is randomly sampled from a range (see Box. 6b). The *scale* parameter represents a measure of the characteristic time scale of a fire regime, whereas the *shape* parameter connects the “hazard of burning” with stand age, with *shape* > 1 denoting an increase in hazard of burning with stand age. No fire events are assigned for the first 3 years after afforestation, after which fires are assigned every year according to a binomial probability of a fire event happening in the current year given the previous fire event and difference between the years. The base probability of fire happening on any given year is set to 1/FRI.

$$f(x;\lambda,k) = \begin{cases} \frac{k}{\lambda}\left(\frac{x}{\lambda}\right)^{k-1} e^{-(x/\lambda)^k}, & x \geq 0, \\ 0, & x < 0, \end{cases}$$

Equation 1: Weibull distribution with shape parameter k and scale parameter λ.

**10)** The fire fraction is set to the historical fire fraction of the cell (see supplementary Fig. 14). The fire regions are assigned for successive years (if more than 1 event) by starting from the bottom of the inventory region and assigning non-overlapping regions with lat range = fire fraction*lat range of inventory region. **11)** The rest of the parameters in the GCBM are set to default including site quality and planting density.

**a)**

| Eco-zone | Soil Type |
|---|---|
| Taiga Plains | Luvisolic (W. Canada), Cryosolic |
| Taiga Shield West | Brunisolic, Cryosolic |

**b)**

| Parameter | Range/List |
|---|---|
| Dominant soil probability (dom_soil_prob) | [0.6, 1, 0.05] |
| Weibull shape | [1.05, 1.61, 0.05][32] |

**c)**

| Eco-zone | Species | Percentage |
|---|---|---|
| Taiga Plains | Black spruce | 81 |
| | Trembling aspen | 13 |
| | Lodgepole pine | 16 |
| Taiga Shield West | Black spruce | 73 |
| | Lodgepole pine | 27 |

**d)**

| Eco-zone | Weibull Scale Parameter |
|---|---|
| Taiga Plains | [100, 200, 5] |
| Taiga Shield West | [500, 700, 5] |

**e)**

| Species Group | Species |
|---|---|
| 0 | Not stocked |
| 1 | ABIE.AMA, PSEU.MEN |
| 2 | CHAM.NOO, THUJ.PLI, TSUG.CAN, TSUG.HET, TSUG.MER |
| 3 | PINU.ALB, PINU.BAN, PINU.CON, PINU.PON, PINU.RES, PINU.STR |
| 4 | PICE.ABI, PICE.ENG, PICE.GLA, PICE.MAR, PICE.RUB, PICE.SIT |
| 5 | ABIE.BAL, ABIE.LAS |
| 6 | BETU.PAP, POPU.BAL, POPU.GRA, POPU.TRE |
| 7 | ACER.MAC, ACER.RUB, ACER, SAH, ALNU.INC, ALNU.RUB, BETU.ALL, FRAX.NIG, LARI.LAR, LARI.OCC, QUER.RUB, ULMU.AME |

**f)**

| Species Group | $a_0$ | $a_1$ | $a_2$ | $a_3$ | $a_4$ | $a_5$ | $a_6$ |
|---|---|---|---|---|---|---|---|
| 0 | 0 | 0 | 0 | 0 | 0 | 0 | 0 |
| 1 | 5.7 | 0.063 | -0.000 | -173.8 | 14.529 | 0 | 1.0423 |
| 2 | 5.7 | 0.063 | -0.000 | -173.8 | 14.529 | 0 | 1.0423 |
| 3 | 6.47 | 0.127 | -0.000 | -67.49 | 2.6486 | 0.011 | 1.0777 |
| 4 | 6.44 | 0.098 | 0.001 | -37.40 | 7.5551 | -0.036 | 1.2127 |
| 5 | 4.84 | 0 | 0.000 | -74.89 | 4.3921 | 0 | 1.2453 |
| 6 | 6.63 | 0 | -0.000 | -55.66 | 0.8537 | 0 | 1.0289 |
| 7 | 6.60 | 0 | -0.000 | -47.11 | 1.6253 | 0 | 1.0819 |

Box 6: **Simulation parameter values and ranges. a)** Dominant soil type mapping according to eco-zone from literature and Soil Landscapes of Canada (SLC v3.2). **b)** Simulation-related parameters and their assumed ranges. [A, B, C] refers to a list from A to B, including A and B, in steps of C. **c)** Species and species percentages according to eco-zone based on NTEMS-SBFI [20]. The smallest percentage is rounded up to include all other smaller species percentages. **d)** Weibull scale parameter ranges according to eco-zones [32]. [A, B, C] refers to a list from A to B, including A and B, in steps of C. **e)** Species assignment to species groups [94]. **f)** Coefficients of yield equation (Eq. 2) [48] according to species groups [94].

## Generating yield curves

$$\text{Yield} = \exp\left(a_0 + a_1 \cdot \text{MAT} + a_2 \cdot \text{PCP} + \frac{a_3 + a_4 \cdot \text{MAT} + a_5 \cdot \text{PCP}}{T_s}\right) \cdot a_6$$

Equation 2: Yield equation derived from Ung et al. [48].

We use two sets of yield curves. The first set of  yield curves are generated according to the equations given in McKenney et al. [94] for different eco-zones and different species groups (groups 0-7). These equations are derived from Ung et al. [48] and are obtained by fitting growth and yield models to temporary sample plots across Canada. As these curves are parametrized using climate variables and were shown to be applicable across provinces, we employ them in our study to study overall trends and consider them as a lower bound. Our region of interest spans multiple provinces and ecoregions, and as these yield curves have been fitted to diverse climates across Canada, they are appropriate for an exploratory study such as this. The general form of the equation is as given in Eq. 2. The species are assigned to species groups as in Box. 6e. The coefficients for the yield equations for different species groups are as given in Box. 6f. The MAT and PCP values used are averages for the eco-zones. The second set of yield curves were obtained by contacting the forestry department of the government of northwest territories (GNWT). The yield curves shared by GNWT are the ones used in the forest management agreement and 25 year strategic plan document prepared by Timberworks Inc in support of their land use permit application [62]. These yield curves are meant to be used in the province of Northwest Territories, and comprise of 5 species types (Deciduous, Black Spruce, White Spruce, Mixed Wood, Pine), 2 density configurations (High, Low), and 3 site qualities (Good, Medium, Poor). In subsequent phases, province specific yield curves will be explored [44-47]. Both these yield curves, other data, and project code can be found under the data folder in the GitHub Repository [95].

It is important to note that GCBM does not dynamically model the full hydrology and energy fluxes of the forest ecosystem. Therefore, using MAT and PCP as inputs for yield curves serves as a surrogate for the complex water-energy nexus that truly governs tree growth and mortality. To get a more accurate estimation of these interactions, combined or coupled modeling approaches are necessary. Future research could involve integrating GCBM's detailed carbon budget accounting with other specialized models, such as a dedicated hydrological model to better simulate soil moisture and water stress, or a land surface model to explicitly calculate energy fluxes and albedo feedbacks. This integrated approach would provide a more holistic understanding of how afforestation impacts, and is impacted by, the complete water and energy balance of the region.

## Research Framework and Simulation Workflow

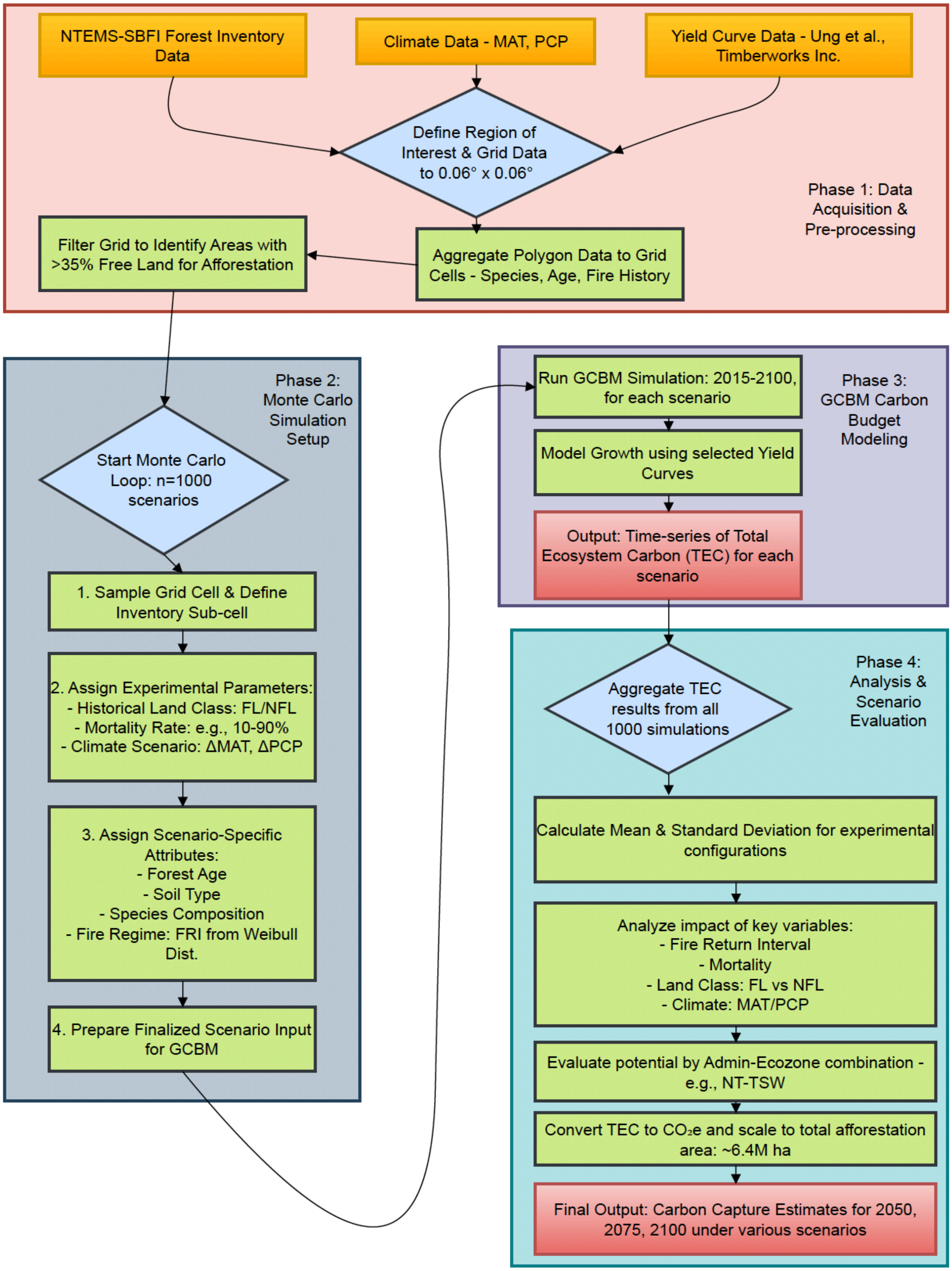

NTEMS-SBFI Forest Inventory Data
Climate Data - MAT, PCP
Yield Curve Data - Ung et al., Timberworks Inc.
Define Region of Interest & Grid Data to 0.06° x 0.06°
Phase 1: Data Acquisition & Pre-processing
Filter Grid to Identify Areas with >35% Free Land for Afforestation
Aggregate Polygon Data to Grid Cells - Species, Age, Fire History
Phase 2: Monte Carlo Simulation Setup
Start Monte Carlo Loop: n=1000 scenarios
1. Sample Grid Cell & Define Inventory Sub-cell
2. Assign Experimental Parameters:
- Historical Land Class: FL/NFL
- Mortality Rate: e.g., 10-90%
- Climate Scenario: ΔMAT, ΔPCP
3. Assign Scenario-Specific Attributes:
- Forest Age
- Soil Type
- Species Composition
- Fire Regime: FRI from Weibull Dist.
4. Prepare Finalized Scenario Input for GCBM
Run GCBM Simulation: 2015-2100, for each scenario
Phase 3: GCBM Carbon Budget Modeling
Model Growth using selected Yield Curves
Output: Time-series of Total Ecosystem Carbon (TEC) for each scenario
Phase 4: Analysis & Scenario Evaluation
Aggregate TEC results from all 1000 simulations
Calculate Mean & Standard Deviation for experimental configurations
Analyze impact of key variables:
- Fire Return Interval
- Mortality
- Land Class: FL vs NFL
- Climate: MAT/PCP
Evaluate potential by Admin-Ecozone combination - e.g., NT-TSW
Convert TEC to CO₂e and scale to total afforestation area: ~6.4M ha
Final Output: Carbon Capture Estimates for 2050, 2075, 2100 under various scenarios

**Box 7: Research Framework and Simulation Workflow.** Outlines the research framework used to estimate the carbon removal capacity of Taiga afforestation. The workflow is divided into four main phases: Data Acquisition and Pre-processing, Monte Carlo Simulation Setup, GCBM Carbon Budget Modeling, and Analysis and Scenario Evaluation.

To provide a clear and logical overview of our study, this section outlines the research framework used to estimate the carbon removal capacity of taiga afforestation. The workflow is divided into four main phases: (1) Data Acquisition and Pre-processing, (2) Monte Carlo Simulation Setup, (3) GCBM Carbon Budget Modeling, and (4) Analysis and Scenario Evaluation. The process is designed to integrate spatially explicit inventory data, probabilistic disturbance regimes, and climatic scenarios to generate robust estimates of carbon dynamics. The logical flow of this framework is illustrated in the flowchart given in Box 7.

**Phase 1: Data Acquisition & Pre-processing**

The initial phase focused on gathering and preparing the necessary spatial and tabular data. We selected the north-western edge of Canada's boreal forest as our region of interest from the NTEMS-SBFI [20]. This region spans the Taiga Plains (TP) and Taiga Shield West (TSW) ecozones. The raw polygon data from NTEMS-SBFI was gridded to a 0.06 x 0.06 degree resolution. For each grid cell, key attributes such as tree species percentages, forest age, and historical fire data were aggregated and averaged. Climate data, specifically mean annual temperature (MAT) and total annual precipitation (PCP), were downloaded from ClimateDataCA [93] and assigned to the corresponding grid cells. Finally, the grid was filtered to retain cells with greater than 35% "free area" available for potential afforestation, creating a baseline inventory for our simulations. See subsections “Region of Interest” and “Compiling the NTEMS-SBFI data in the region of interest” in the “Methods” section for more details.

**Phase 2: Monte Carlo Simulation Setup**

To account for variability and uncertainty, we employed a Monte Carlo approach, creating 1000 independent simulations for each experimental scenario configuration. Each scenario was built by randomly sampling a grid cell from the pre-processed inventory. Within each scenario, key parameters were assigned based on the specific experiment being run. This included defining the historical land class (historically forested 'FL' or non-forested 'NFL'), assigning forest age, and selecting species for afforestation based on ecozone-specific data. A critical component of this phase was the assignment of fire events. Fire return intervals (FRIs) were sampled from ecozone-specific Weibull distributions, which effectively capture the observed fire regimes in the boreal region. Other parameters, such as generic mortality rates and the timing of afforestation (set to 2025), were also defined at this stage. See subsection “Creating scenarios” in the “Methods” section for more details.

**Phase 3: GCBM Carbon Budget Modeling**

With the scenarios fully parameterized, we used the GCBM to simulate forest carbon dynamics. The GCBM simulates changes in all five IPCC carbon pools: aboveground biomass, belowground biomass, litter, dead wood, and soil organic carbon. For each of the 1000 scenarios, we ran a simulation from the start year of 2015 to 2100. Growth and yield were modeled using two sets of yield curves: a conservative set derived from Ung et al. and a more optimistic set from Timberworks Inc. based on Alberta's GYPSY model. The model's output consisted of time-series data on the total ecosystem carbon (TEC) for each simulated patch. See subsections “Generic Carbon Budget Model (GCBM)” and “Generating yield curves” in the “Methods” section for more details.

**Phase 4: Analysis & Scenario Evaluation**

In the final phase, the TEC outputs from all simulations were aggregated and analyzed. We calculated the mean and standard deviation of carbon removal across various scenarios to understand the influence of key variables like FRI, mortality, historical land class, and climate parameters (MAT and PCP). The results were evaluated for different administrative-ecozone combinations (e.g., NT-TSW) to identify regions with the highest sequestration potential. Finally, we converted the TEC (tonnes of Carbon) into tonnes of $CO_2$ equivalent ($CO_2e$) and scaled the results to the total available afforestation area (~6.4 M hectares) to provide comprehensive estimates of carbon removal potential by 2050, 2075, and 2100 under different illustrative scenarios.

## Forest Inventory of the Region of Interest

The region of interest at the northwestern boreal edge exhibits distinct spatial patterns in its forest characteristics (Box. 8). For clarity and to better illustrate the spatial nuances of the inventory data, we focus specifically on the Northwest Territories (NT). This region was chosen because it is the largest administrative zone by area in our study and our results indicate it has the highest carbon sequestration potential. Focusing on the NT provides a detailed and representative view of the key environmental gradients and forest characteristics without the visual clutter of showing all administrative regions at once. The compiled, gridded inventory data has been made available in the project's GitHub repository to allow for replication and to serve as a resource for other researchers.

The forest age distribution reveals that older stands, with ages exceeding 75 years, are predominantly located in the southern and central parts of the study area, while younger forests are scattered throughout (Box. 8a). The percentage of free area, available for potential afforestation, is highest in the northern parts of the region, often exceeding 70%, which corresponds to the transition into the taiga and tundra ecosystems (Box. 8b). Conversely, the percentage of forested area is densest in the southern portion, with coverage frequently greater than 30%, and gradually decreases northward (Box. 8c). A clear latitudinal gradient is observed for mean annual temperature, with warmer temperatures (around -2.1°C) in the south and progressively colder conditions toward the north, reaching as low as -9.9°C (Box. 8d). The dominant tree species across the region is Black Spruce (*Picea mariana*), which covers the majority of the landscape. Pockets of Jack Pine (*Pinus banksiana*), Tamarack Larch (*Larix laricina*), and White Spruce (*Picea glauca*) are also present, with minor occurrences of other species like Trembling Aspen (*Populus tremuloides*) and Paper Birch (*Betula papyrifera*) (Box. 8e).

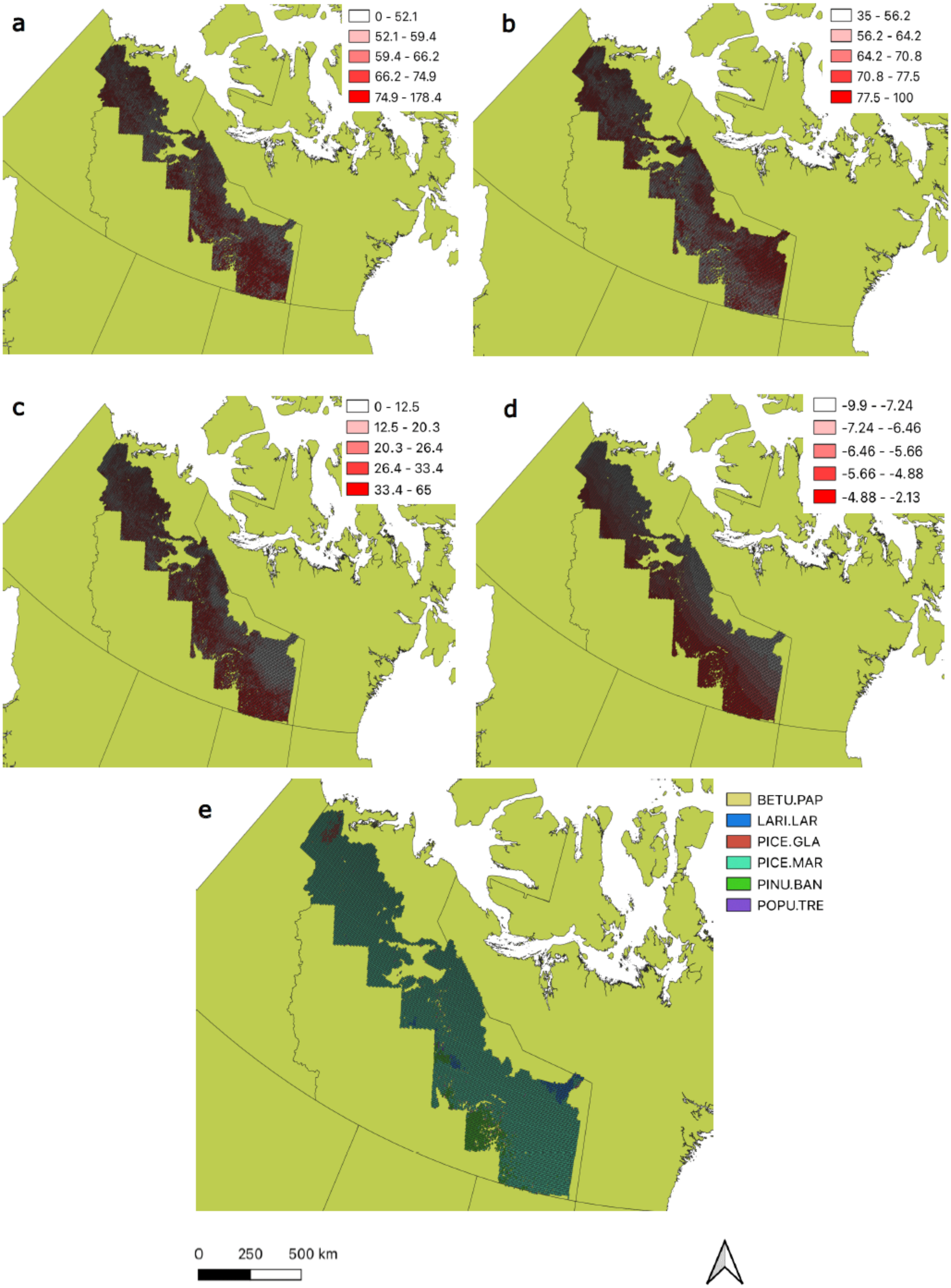
a
0 - 52.1
52.1 - 59.4
59.4 - 66.2
66.2 - 74.9
74.9 - 178.4
b
35 - 56.2
56.2 - 64.2
64.2 - 70.8
70.8 - 77.5
77.5 - 100
c
0 - 12.5
12.5 - 20.3
20.3 - 26.4
26.4 - 33.4
33.4 - 65
d
-9.9 - -7.24
-7.24 - -6.46
-6.46 - -5.66
-5.66 - -4.88
-4.88 - -2.13
e
BETU.PAP
LARI.LAR
PICE.GLA
PICE.MAR
PINU.BAN
POPU.TRE
0
250
500 km

Box. 8: **Forest inventory characteristics for the Northwest Territories portion of the study area.** The data is compiled from the NTEMS-SBFI and gridded. Plotted variables include: **a)** forest age in years, **b)** percentage of free area available for afforestation, **c)** percentage of existing forest cover, **d)** mean annual temperature in degrees Celsius, and **e)** dominant tree species. Species abbreviations are as follows: BETU.PAP (*Betula papyrifera*), LARI.LAR (*Larix laricina*), PICE.GLA (*Picea glauca*), PICE.MAR (*Picea mariana*), PINU.BAN (*Pinus banksiana*), and POPU.TRE (*Populus tremuloides*). Note: Each cell can contain a list of species according to their observed percentages. Only the most dominant species is visualized here.

## Validation Methodology

To assess the credibility of the GCBM projections, we conducted validation against plot-level data from Canada's National Forest Inventory (NFI) [67]. Data processing and variable definitions followed the official NFI compilation procedures [96] and data dictionaries [97]. The NFI ground plots were selected as the primary validation dataset due to their systematic national-scale design and comprehensive carbon pool measurements. This validation was designed to test the model's performance across a range of stand structures and developmental stages found within the study area.

### Primary Validation Dataset and Processing

Our validation relies on data from NFI ground plots located within the Northwest Territories portion of the Taiga Plains and Taiga Shield West ecozones. All complete plot visits from the available measurement cycles were processed to derive carbon stocks for five key ecosystem pools.

### Live Biomass Carbon (AGB + BGB)

The live tree carbon pool was calculated from individual tree measurements within the NFI database. Plot-level Above-Ground Biomass (AGB) (Tonnes/ha) was calculated by summing the per-stem biomass values provided in the NFI tree tables (*biomass_total* for large trees, *smtree_biomass* for small trees) and scaling the total by the plot area. Below-Ground Biomass (BGB) was not directly measured but was estimated by applying standard root-to-shoot ratios to the AGB of each plot. Based on national defaults, a ratio of 0.20 was used for conifer-dominated plots and 0.26 for broadleaf-dominated plots. Live Carbon was calculated by summing AGB and BGB and applying a carbon fraction of 0.5.

### Woody Debris Carbon

The carbon stock in the dead wood pool was calculated directly from the *nt_gp_wd_summary.csv* file. The values from the four component columns (*plotbio_swd, plotbio_wd, plotbio_roundwd, plotbio_oddwd*), which represent biomass in tonnes per hectare, were summed and multiplied by a carbon fraction of 0.5 to yield total woody debris carbon in Tonnes C/ha.

### Soil Organic Carbon (SOC)

The soil organic carbon pool was calculated to the full depth of measured samples by combining data from the three NFI soil sample files: *nt_gp_for_flr_org_sample.csv*, *nt_gp_soil_org_sample.csv*, and *nt_gp_soil_mineral_sample.csv*. The calculation did not require the horizon description file and instead relied on the depth measurements recorded with each individual sample. Layer Thickness, the thickness of each soil layer was determined directly from the *sample_upper* and *sample_bottom* columns in the sample files. To handle inconsistencies across files, a unified data-selection logic was implemented. For organic layers, carbon content was derived from the *tc_8mm* column. For mineral layers, the *tc* column was used. These were converted from g/kg to percent by dividing by 10. For bulk density, *bulk_density_total* was prioritized, with *bulk_density_2mm* used as a fallback. To ensure physical realism, a data quality filter was applied, removing any soil sample records with a calculated bulk density outside the range of 0.1–1.5 g/cm³ or a carbon content outside the range of 0–60%. The carbon stock for each layer (in Tonnes C/ha) was calculated by multiplying its thickness (m), bulk density (kg/m$^3$), and carbon fraction, adjusted by the appropriate unit conversion factor. These values were then summed to the plot level.

**Total Ecosystem Carbon (TEC)**
The total ecosystem carbon stock for each plot-visit was calculated as the sum of the live carbon (AGB + BGB), woody debris carbon, and soil organic carbon pools.

**Stratification for Model Comparison**
To facilitate a direct comparison with GCBM outputs, the processed NFI plot data was stratified based on three key stand attributes:

1. **Stand Age**: Plots were grouped into 20-year age classes (e.g., 0-20, 20-40 years).
2. **Density Class**: Following our yield curve parameterization, plots were classified as either Sparse (<900 live stems ≥9 cm DBH per hectare) or Dense (≥900 stems/ha).
3. **Leading Species**: The dominant species for each plot was identified as the one with the highest percentage contribution in the NFI's species composition tables.

**Validation Approach and Performance Metrics**
The validation was conducted in two distinct parts to robustly assess different aspects of model performance:

1. **Stock Validation**: The primary validation was performed on carbon stocks (Tonnes C/ha). The mean (from *n* NFI plots) of calculated carbon pools (Live C, TEC) was compared against the corresponding mean stock generated by the *n* GCBM outputs for each stratum. This approach provides an assessment of the model's ability to represent the state of the ecosystem at different developmental stages.
2. **Increment Validation**: The analysis of annual carbon increments (Tonnes C/ha/yr) was confined to the Live Carbon pool only. While soil depth was harmonized between visits to ensure a valid comparison, the resulting high variance in the soil and deadwood increment data confirmed that the measurement uncertainty over short NFI re-measurement intervals is too large to provide a reliable signal for validating these slow-turnover pools. The live carbon increment, however, offers a clear and direct measure of forest growth dynamics. To facilitate a direct comparison, GCBM outputs for each stratum were simulated forward for a duration matching the average re-measurement interval of the corresponding NFI plots, from which an annual increment was calculated. The mean annual increment predicted from the *n* GCBM runs was then compared with the mean annual increment from the *n* NFI measurement intervals.

Model performance for both stocks and increments was evaluated using two standard metrics. For each carbon pool, the performance metrics, Coefficient of Determination ($R^2$) and Relative Root Mean Square Error (RMSE) were calculated across the full set of strata, comparing the list of NFI-derived means against the list of GCBM-predicted means. This provides a robust assessment of model performance over the entire range of simulated conditions.

## Rationale for Using Total Ecosystem Carbon (TEC) as the Main Indicator
In this study, Total Ecosystem Carbon (TEC) was selected as the primary indicator for quantifying the climate mitigation potential of afforestation. This choice is deliberate and serves several key purposes:

1. **Comprehensive Scope:** TEC provides a holistic measure of carbon dynamics by accounting for all five ecosystem carbon pools mandated by the Intergovernmental Panel on Climate Change (IPCC): aboveground biomass, belowground biomass, litter, dead wood, and soil organic carbon. This approach avoids a narrow focus on only live biomass, which would overlook significant carbon stocks and fluxes within the dead organic matter and soil pools, pools that are particularly large and vulnerable in boreal ecosystems.
2. **Policy Relevance:** As the scientific foundation of the GCBM is the CBM-CFS3 model used for Canada's national forest carbon reporting, using TEC aligns our study with established national and international reporting standards. This ensures that our findings are directly comparable and can be integrated into broader assessments of how land-use activities contribute to Canada's climate targets.
3. **Disturbance Impact Assessment:** Boreal forests are disturbance-driven ecosystems, with factors like fire playing a powerful role in carbon cycling. TEC is particularly well-suited to capture the net effect of these disturbances, as it reflects not only the loss of carbon from biomass combustion but also the subsequent transfers of carbon to dead organic matter pools and the long-term changes in soil carbon. By tracking TEC, we can more accurately model the ecosystem's response to changing fire regimes and mortality events over the 75-year simulation period.

While other metrics, such as net primary productivity or individual pool fluxes, are valuable for detailed biogeochemical analysis, TEC serves as the most robust and policy-relevant indicator for assessing the overall impact of a large-scale afforestation initiative on the landscape-level carbon balance.

## Data Availability

The data that support the findings of this study are publicly available to download and are referenced in the bibliography. Refer to the Methods section for more details. The data generated from the project can be found in the GitHub Repository [95].

## Code Availability

The code repository for this project, including data processing and simulations can be found in our GitHub Repository [95].

## Acknowledgements

We thank the Natural Sciences and Engineering Research Council of Canada (NSERC) for primarily funding this research through the NSERC Alliance Mission grant - ALLRP 577126-2022 (Y.L., R.B., and J.M.-C.). In addition, we acknowledge support from the Canada Research Chairs Program - CRC-2023-00181 (J.M-.C.) and NSERC PDF program (K.B.D.). We also thank Max Fellows and Stephen Kull from the GCBM team for helping us set up GCBM.

## Authors' contributions

K.B.D. conducted the research, performed simulations, wrote the article, plotted results, and created illustrations. E.O. helped with ideation and literature review. R.B., J.M.-C., and Y.L. were involved in the acquisition of funding, editing the article and overall supervision. All authors contributed to discussion and conceptualization of arguments.

## Competing interests

The authors declare no competing interests

# Supplementary

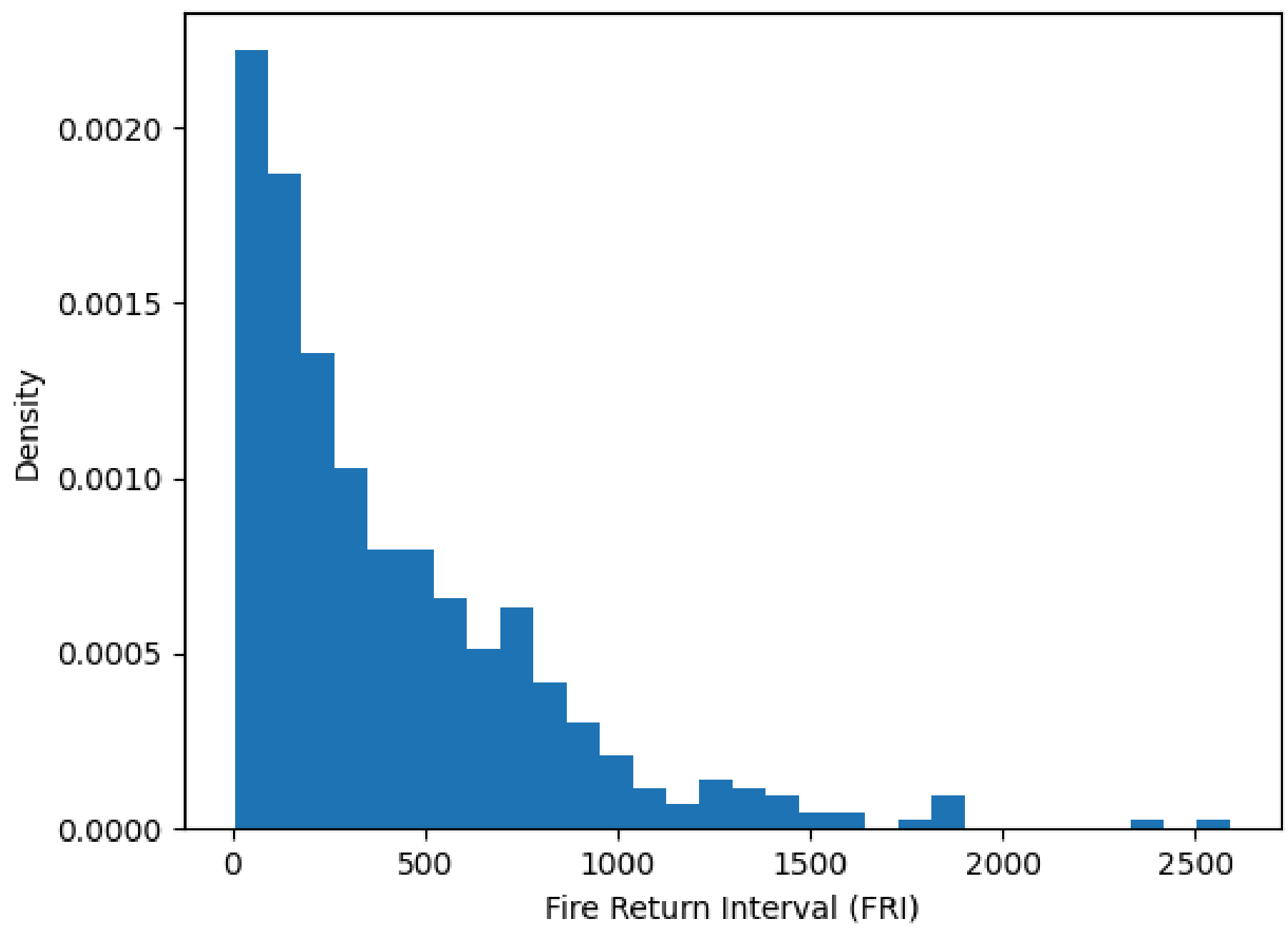

Figure 1: **Histogram of mean fire return interval (FRI)**. A single experiment consists of 1000 independent simulation runs. FRI sampled from a weibull distribution with parameters dependent on eco-zone.

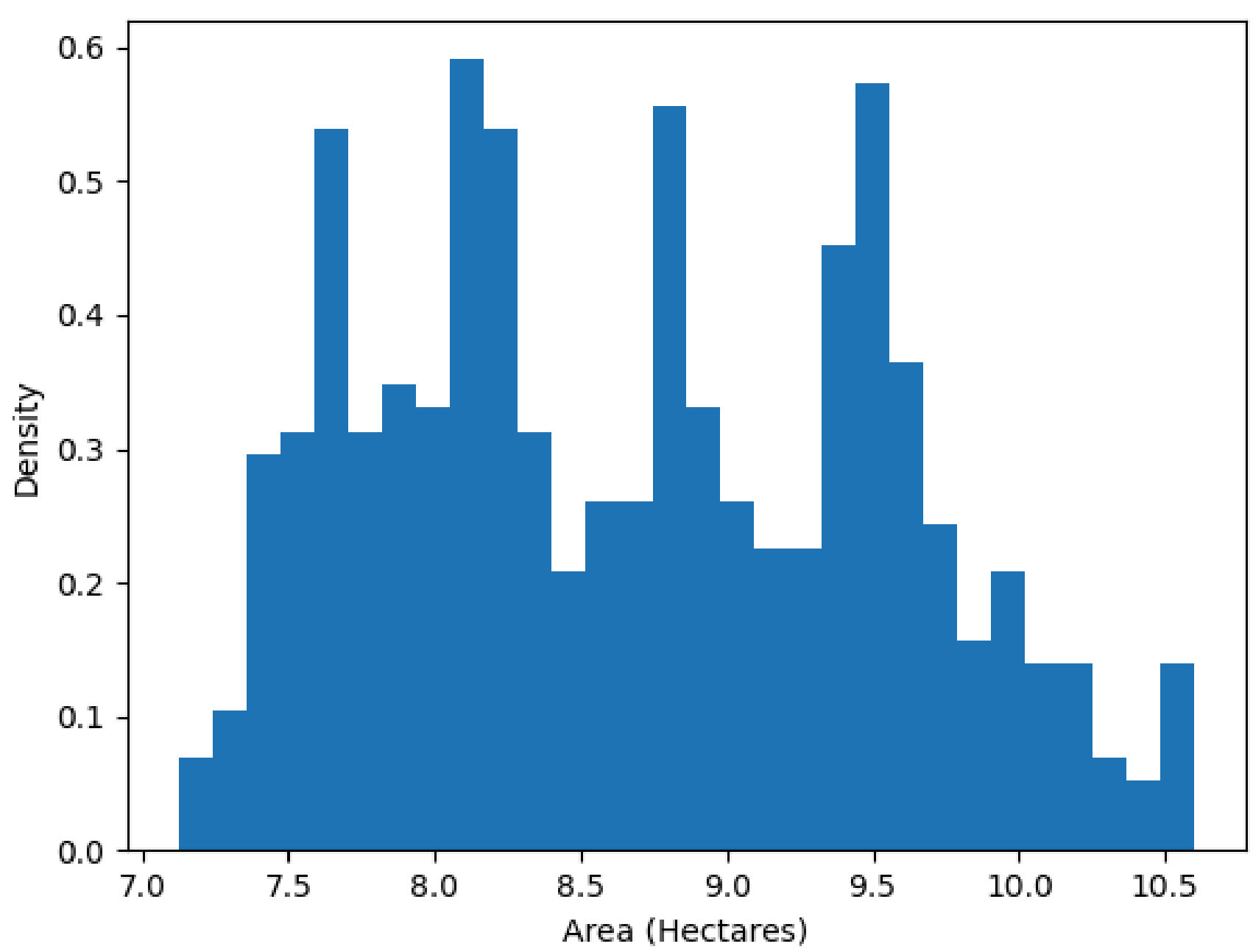

Figure 2: **Histogram of area afforested**. A single experiment consists of 1000 independent simulation runs.

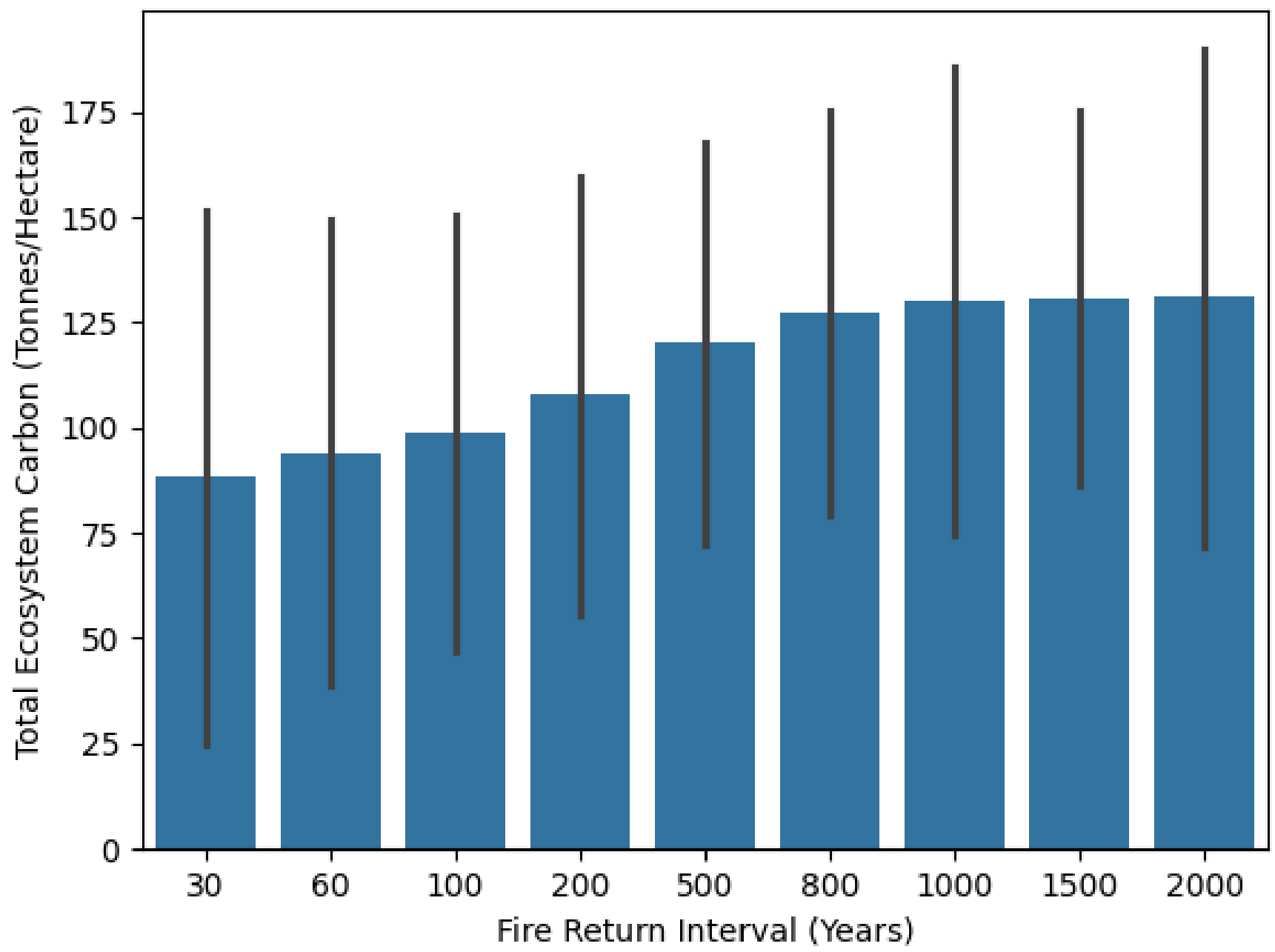

Figure 3: **TEC at 2100 for different mean fire return intervals (FRI) for afforestation on forested land (FL).** The TEC accounts for existing plus reforested trees in FL. The bars denote the means and spread denotes the standard deviations.

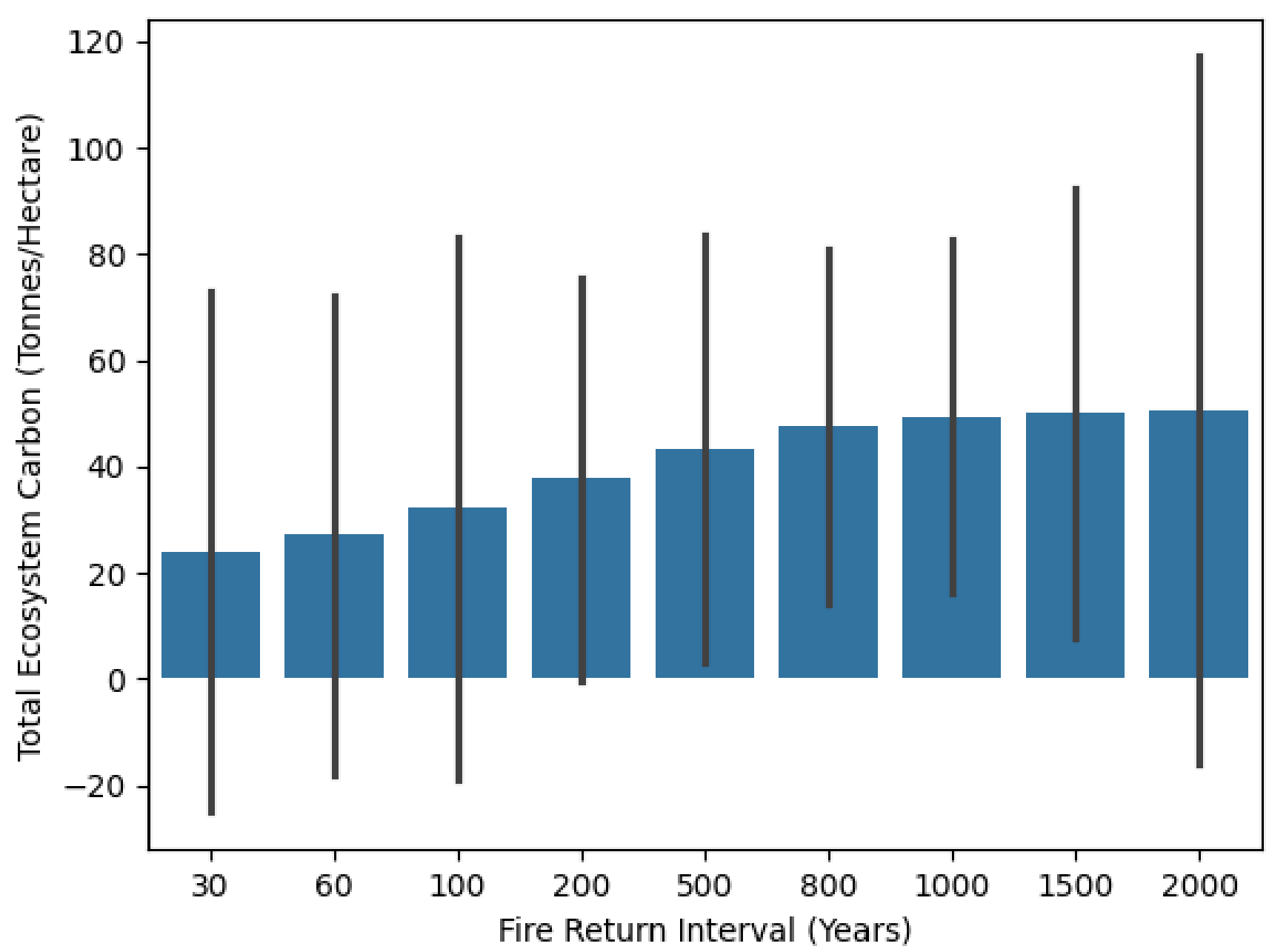

Figure 4: **Total ecosystem carbon (TEC) at 2100 for different mean fire return intervals (FRI) for afforestation on non-forested land (NFL).** The bars denote the means and spread denotes the standard deviations.

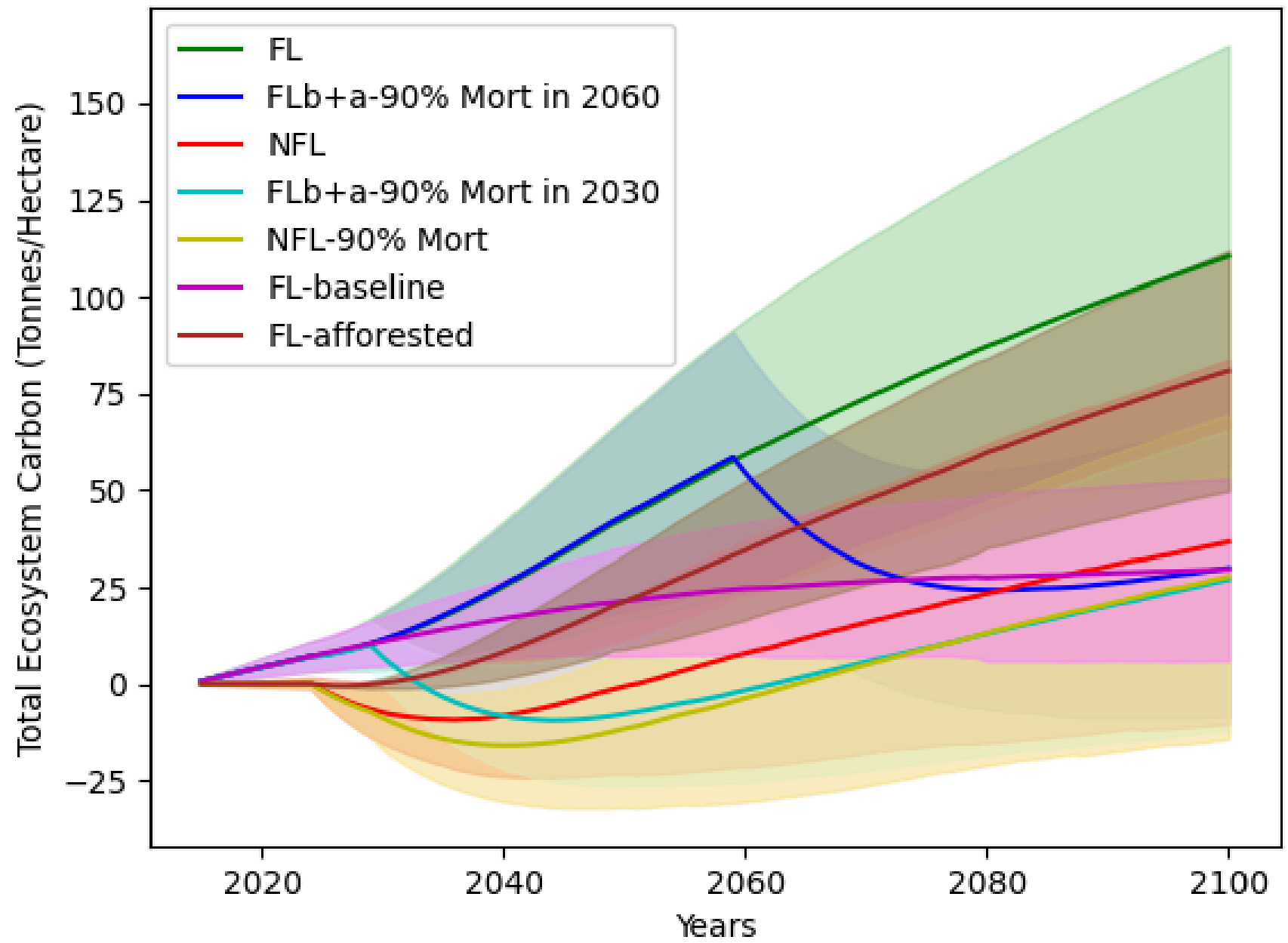

Figure 5: **Standard deviations for main Fig. 3a.** FL - forested land, NFL - non-forested land, X% Mort - X% mortality of afforested trees after 5 years. The baseline refers to existing forests without additional afforestation in free areas. The plotted values are averaged across all other variables. The lines denote the means and the spread shows the standard deviation. .

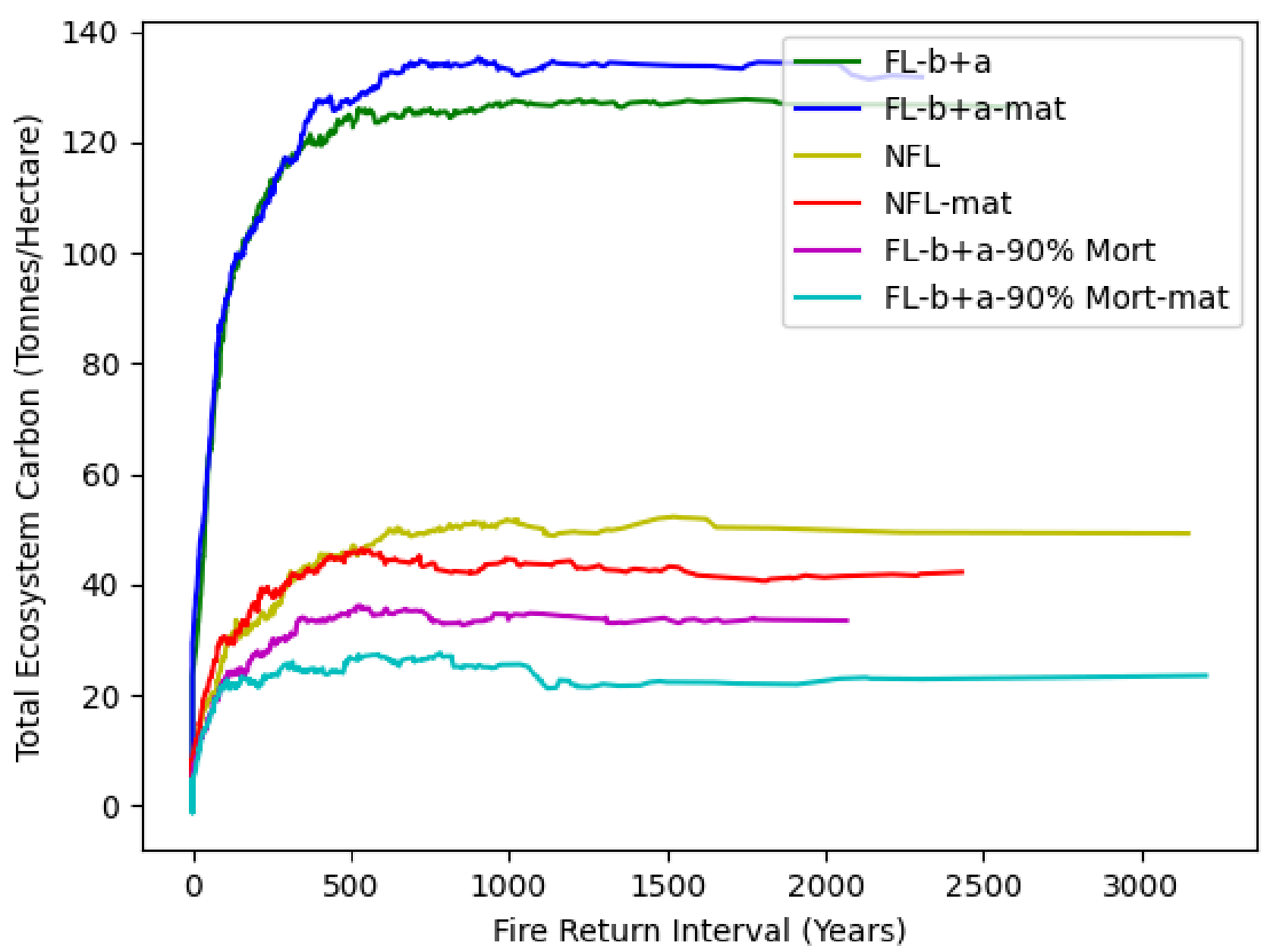

Figure 6: **TEC at 2100 as a function of mean fire return interval (FRI).** FL - forested land, NFL - non-forested land, 90% Mort - 90% mortality of afforested trees after 5 years. Baseline (b) refers to existing forests without additional afforestation in free areas, and (a) to afforestation in free areas. Comparison between when mean annual temperature (MAT) is explicitly passed as input to GCBM versus when it's not. Standard deviation spread not shown to retain clarity.

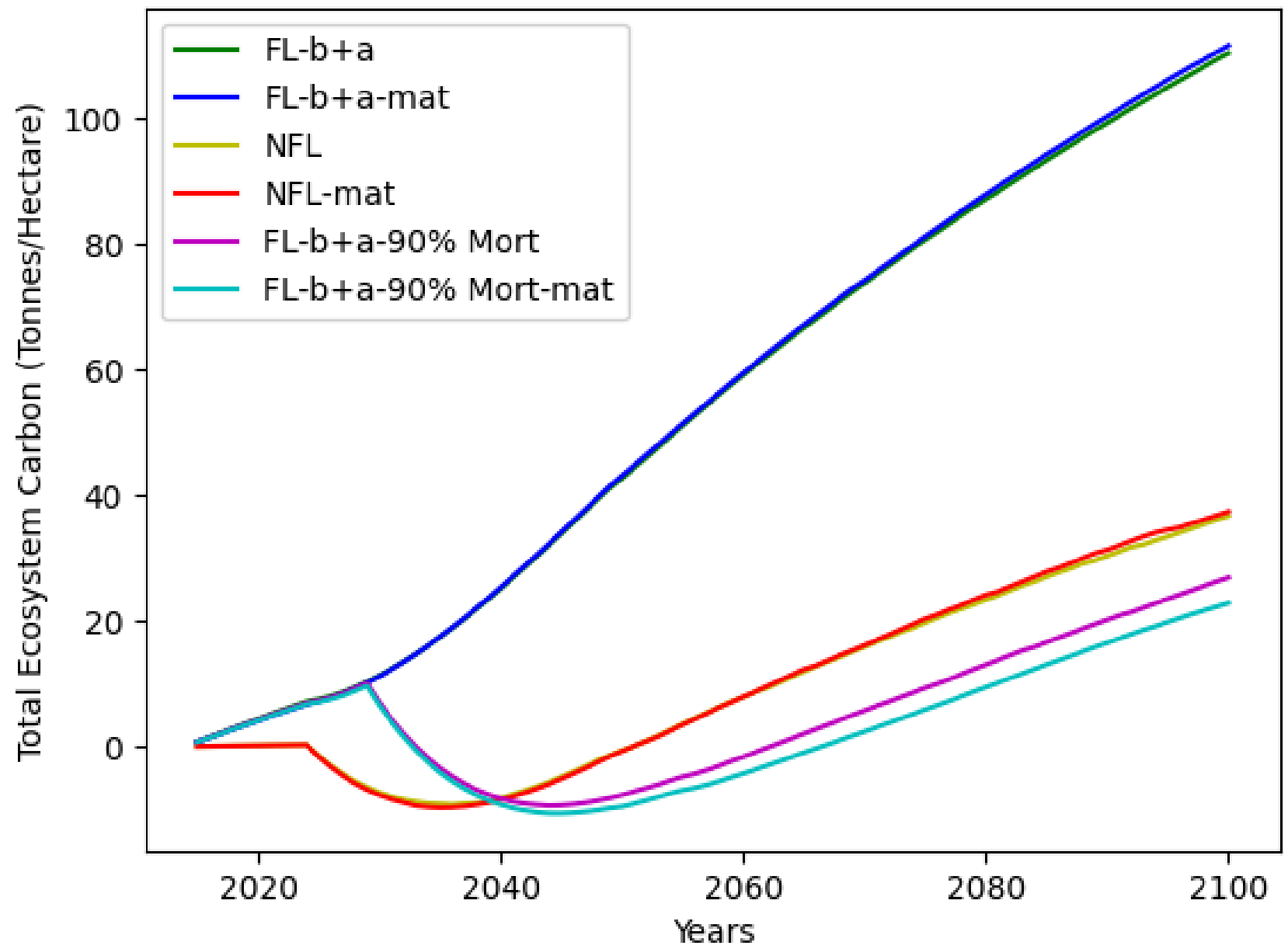

Figure 7: **TEC over the years.** FL - forested land, NFL - non-forested land, 90% Mort - 90% mortality of afforested trees after 5 years. Baseline (b) refers to existing forests without additional afforestation in free areas, and (a) to afforestation in free areas. Comparison between when mean annual temperature (MAT) is explicitly passed as input to GCBM versus when it's not. Standard deviation spread not shown to retain clarity.

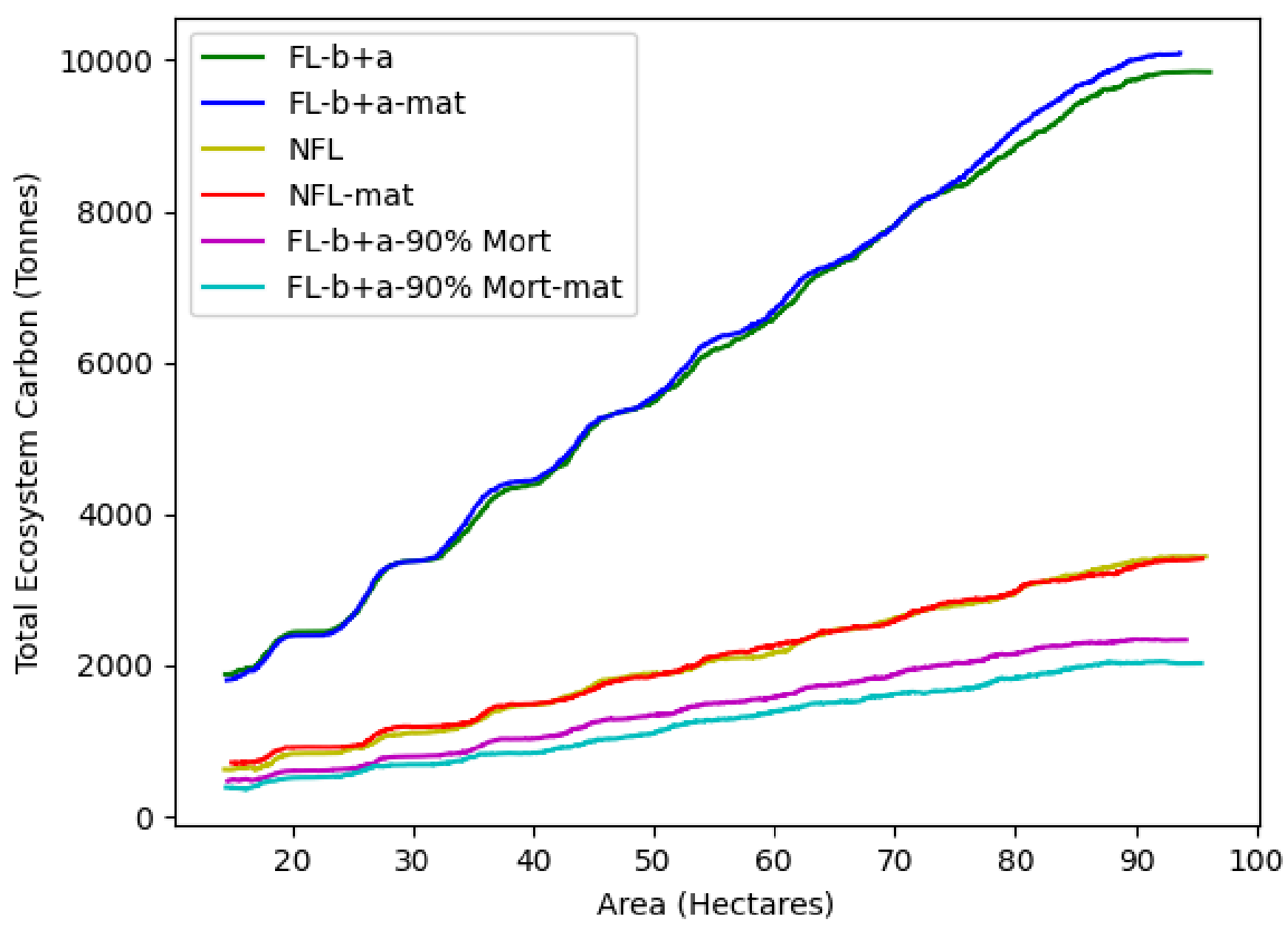

Figure 8: **TEC as a function of the afforestation area.** FL - forested land, NFL - non-forested land, 90% Mort - 90% mortality of afforested trees after 5 years. Baseline (b) refers to existing forests without additional afforestation in free areas, and (a) to afforestation in free areas. Comparison between when mean annual temperature (MAT) is explicitly passed as input to GCBM versus when it's not. Standard deviation spread not shown to retain clarity.

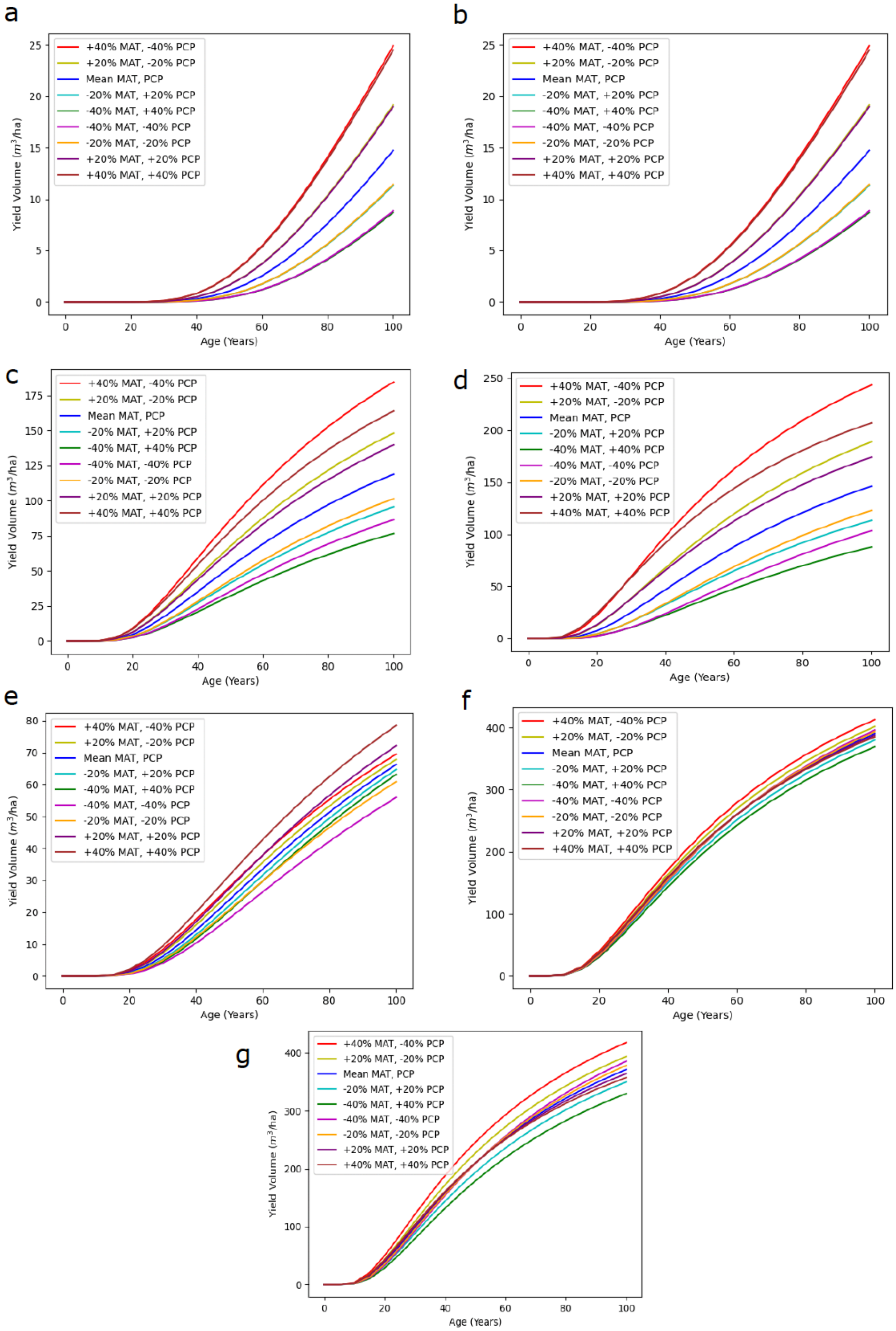

Figure 9: **Yield volume (Ung et al.) as a function of age up to 100 years for different species groups in Taiga Plains eco-zone.** Species groups: a-1, b-2, c-3, d-4, e-5, f-6, g-7. Legend shows varying combination of increase or decrease in mean annual temperature (MAT) and total annual precipitation (PCP).

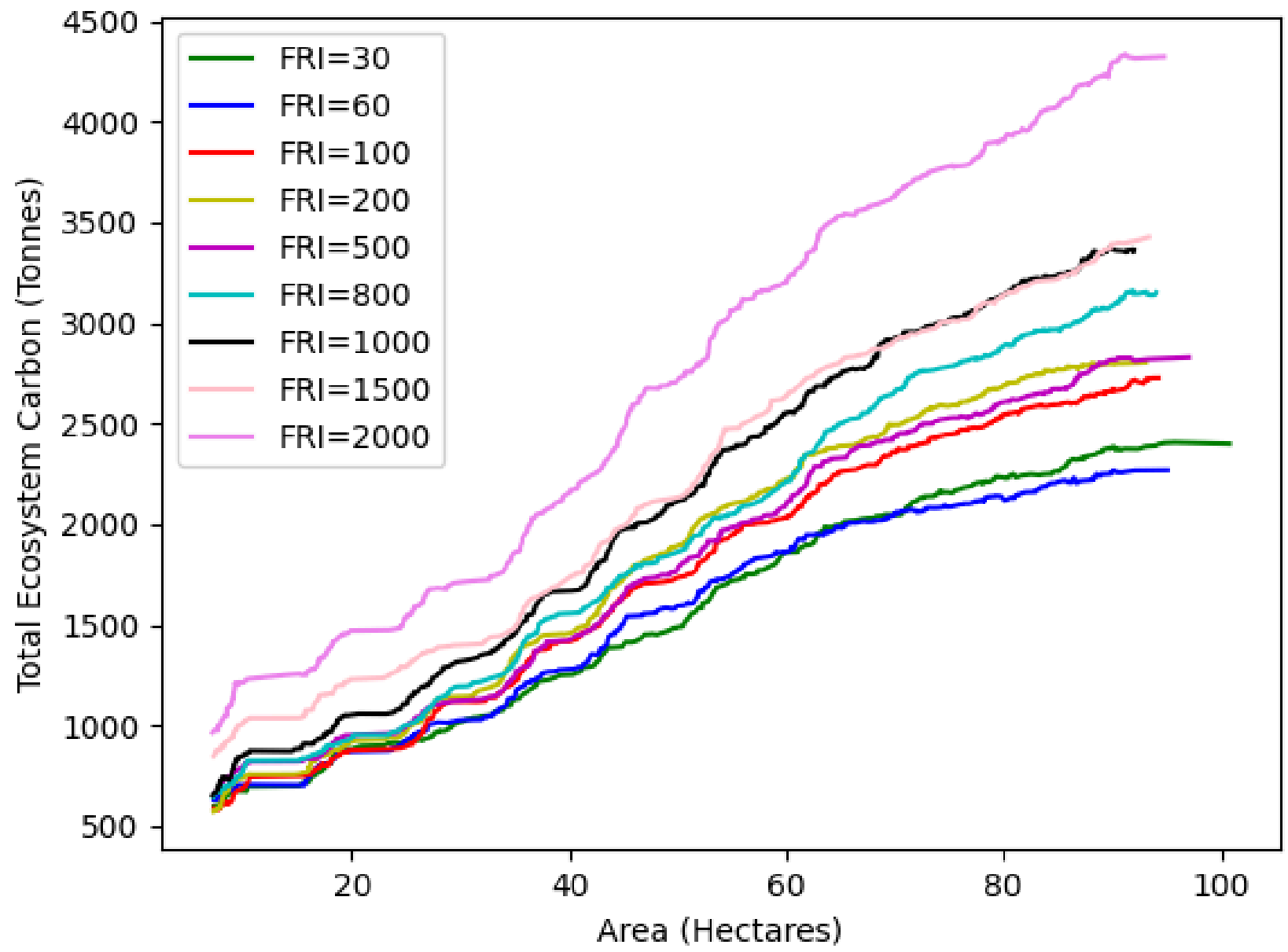

Figure 10: **TEC at 2100 as a function of the afforestation area with different mean fire return intervals (FRI) for afforestation on non-forested land (NFL).** A combination of existing simulations is sampled to get increasing afforestation area and the resulting TEC is added. Standard deviation spread not shown to retain clarity.

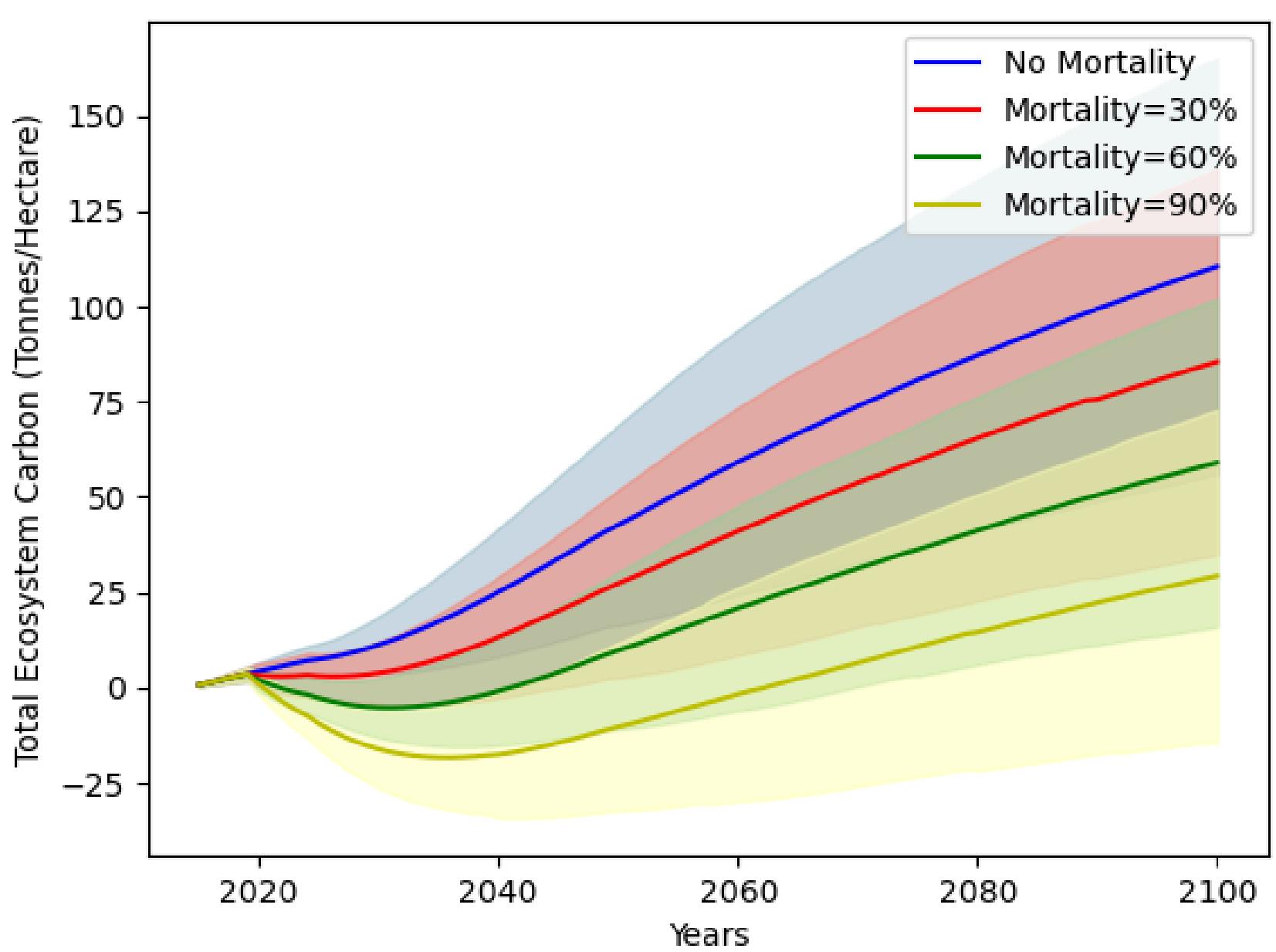

Figure 11: **Standard deviations for main Fig. 7a.** TEC over the years. The lines denote the mean and the spread shows the standard deviation. The TEC accounts for existing plus afforested trees on FL.

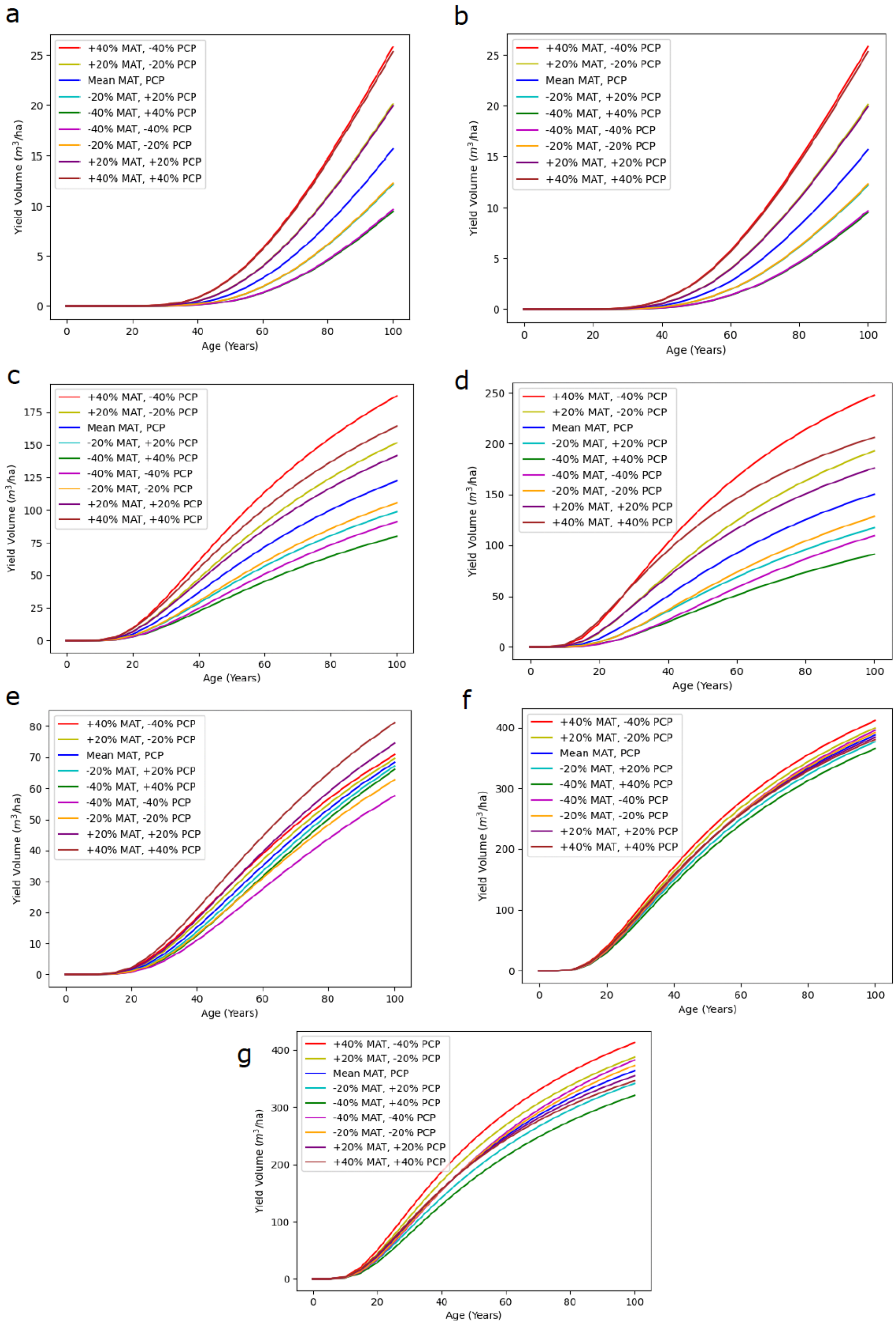

Figure 12: **Yield volume (Ung et al.) as a function of age up to 100 years for different species groups in Taiga Shield West eco-zone.** Species groups: a-1, b-2, c-3, d-4, e-5, f-6, g-7. Legend shows varying combination of increase or decrease in mean annual. temperature (MAT) and total annual precipitation (PCP).

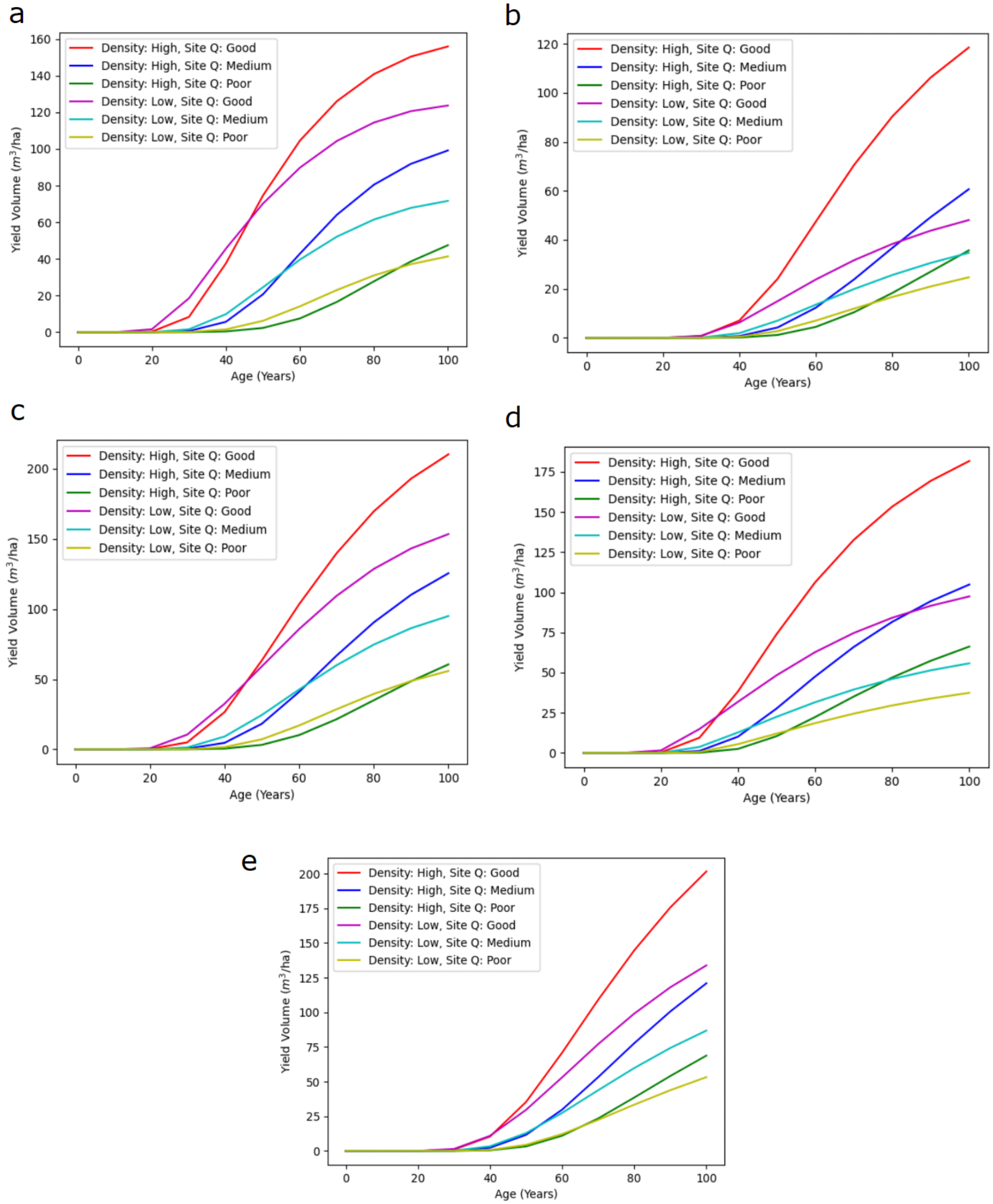

Figure 13: **Yield volume (Timberworks Inc.) as a function of age up to 100 years for different species in Northwest Territories.** Species: a-Deciduous, b-Black Spruce, c-Mixed Wood, d-Pine, e-White Spruce. Legend shows combinations of density and site quality.

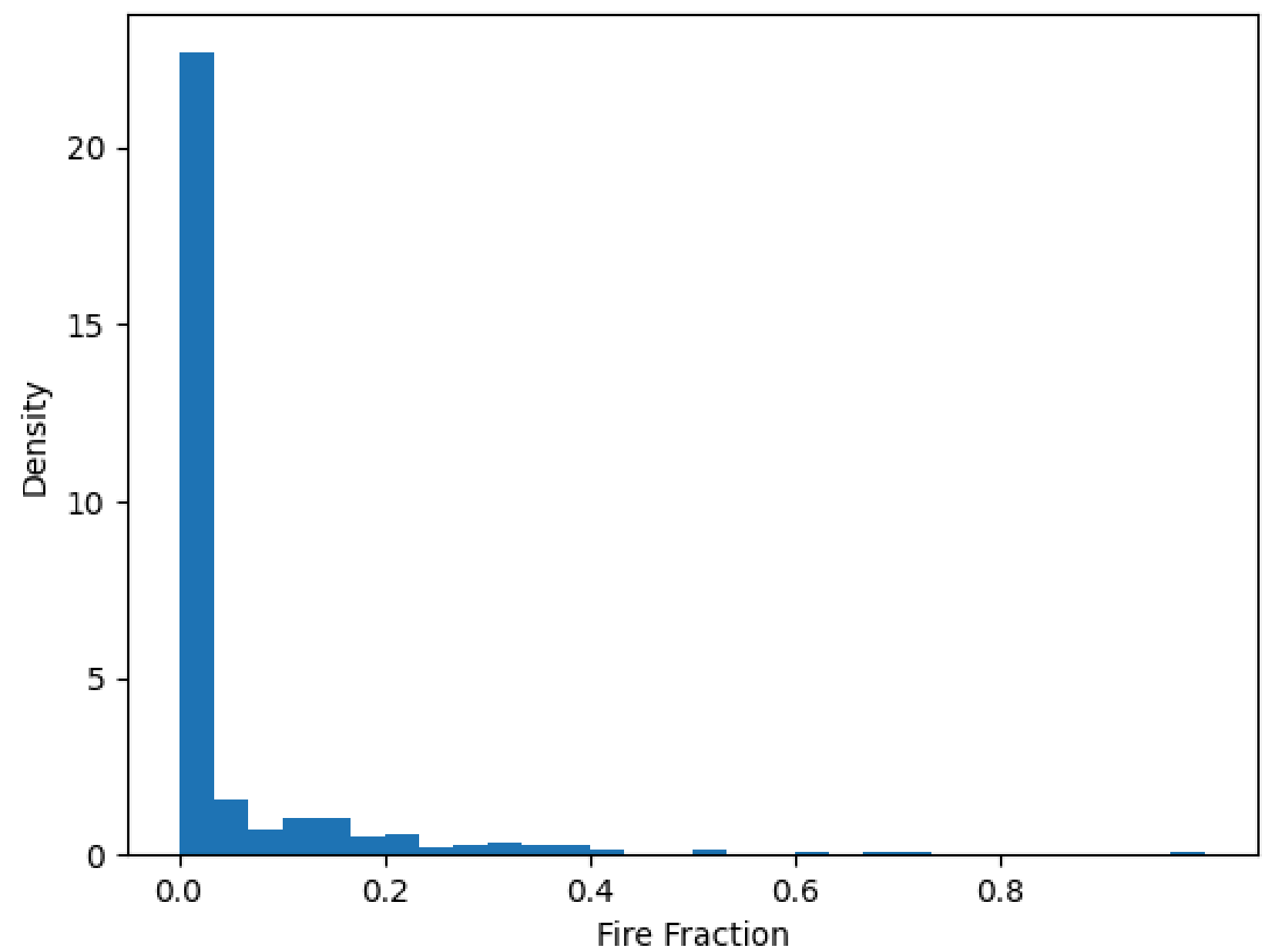

Figure 14: **Histogram of historical fire fractions**. A single experiment consists of 1000 independent simulation runs.

## Ablations with Land Classes, Soil Types, and Fire Regimes

We compare TEC for afforestation on four different historical land classes, forested land (FL), non-forested land (NFL), cropland (CL), and grassland (GL). Here the NFL, CL, GL experiments are not run on patches of land that are classified in those categories according to the land cover classification for Canada, but are assumed to be of that historical land class in cells that have mean forest age zero and percentage free area >=70%. This gives us an idea of what the TEC bounds would look like if the non-forested spaces were made of these land classes in our region of interest. We see that afforestation in NFL, CL, and GL historical land classes roughly results in the same TEC, with NFL>CL>GL marginally (Fig. 15a,b). Compared to these non-forested land classes, afforestation in historically forested land results in substantially better TEC at the end of the century (Fig. 15a,b). Moreover, it shows a much better relationship with the fire regime as the TEC growth makes up for the loss through fire disturbances (Fig. 15a), as well as a much steeper TEC growth over the years (Fig. 15b). This result demonstrates that the TEC dynamics over different historical land classes need to be taken into account before they are chosen for afforestation.

| Eco-zone | Soil Type |
|---|---|
| Taiga Plains | Luvisolic (W. Canada), Cryosolic |
| Taiga Shield West | Brunisolic, Cryosolic |

Table 1: **Dominant soil type mapping according to eco-zone.** Obtained from literature and Soil Landscapes of Canada (SLC v3.2).

To compare the effect of dominant soil types, we run three experiments on FL, one with Cryosolic type removed from Table 1, one with only Cryosolic, and one with randomly choosing from the types in Table 1 (see Fig. 16a). We observe that the soil type configuration doesn't influence the outcome in GCBM, and specificity for soil types might need to be factored into the yield curves that one choses (see Fig. 16a). As we don't have access to yield curves for different soil types, studying this further is deferred to a future study.

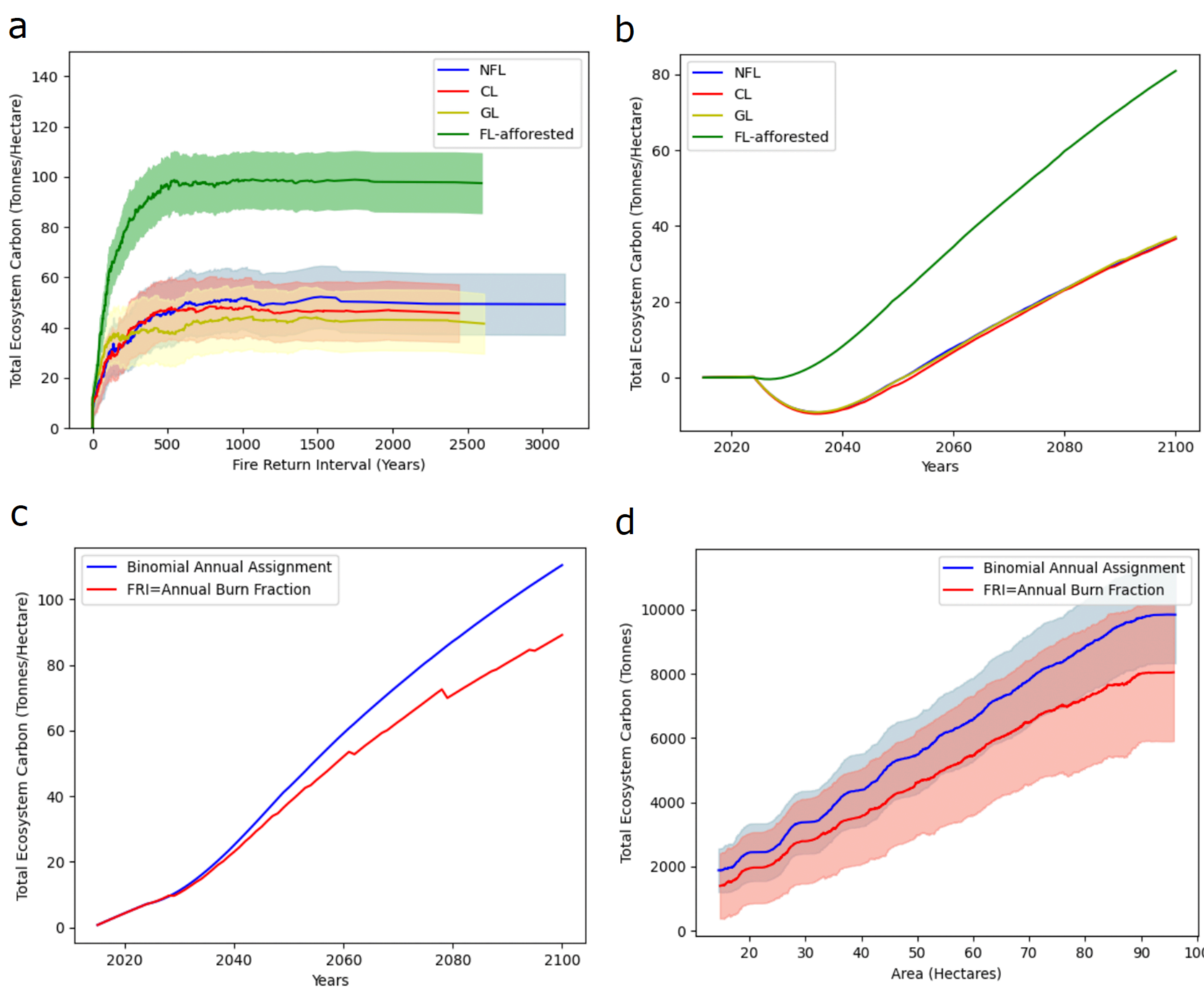

Figure 15: **Ablations with Land Classes, Soil Types, and Fire Regimes. a)** TEC at 2100 as a function of mean fire return interval (FRI) for different historical land classes. FL - forested land, NFL - non-forested land, CL - cropland, GL - grassland. The plotted values are averaged across a window of 100. The lines denote the mean and the spread shows the standard deviation. **b)** TEC over the years up to 2100 for different historical land classes. FL - forested land, NFL - non-forested land, CL - cropland, GL - grassland. The lines denote the means. The standard deviations can be found in Fig. 16b. **c)** TEC over the year for alternate fire regimes on forested land (FL). The TEC accounts for existing plus afforested trees on FL. FRI refers to the fire return interval. The standard deviations can be found in Fig. 16c. **d)** TEC at 2100 as a function of the afforestation area. The TEC accounts for existing plus afforested trees on FL. FRI refers to the fire return interval. A combination of existing simulations is sampled and combined to get higher afforestation area and the resulting TEC is added. The lines denote the mean and the spread shows the standard deviation.

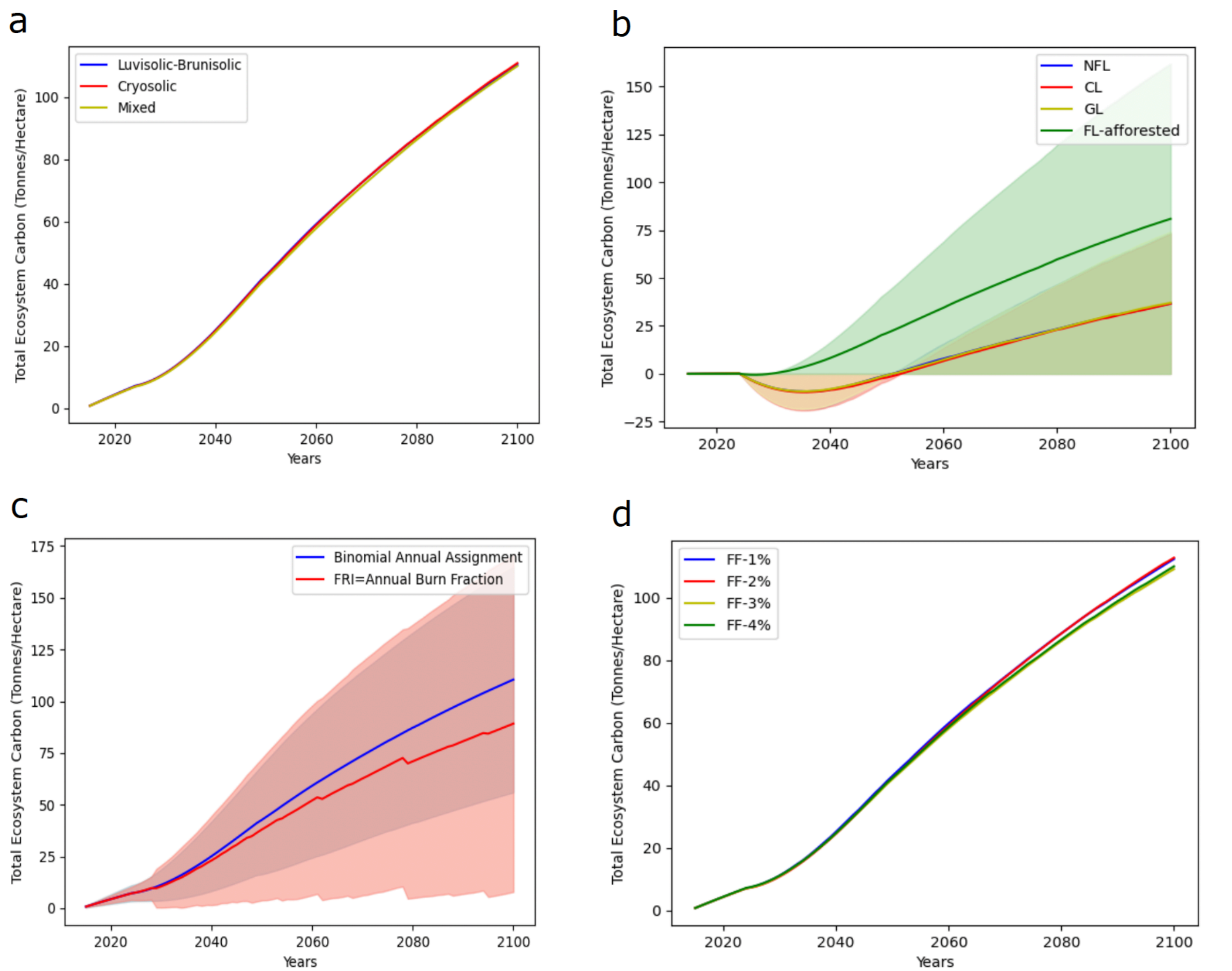

Figure 16: **a)** TEC over the years with different dominant soil types on forested land (FL). The TEC accounts for existing plus afforested trees in FL. Standard deviation spread not shown to retain clarity. **b)** Standard deviations for Fig. 15b. **c)** Standard deviations for Fig. 15c. **d)** TEC over the years with different fire fractions on forested land (FL). The TEC accounts for existing plus afforested trees in FL. Standard deviation spread not shown to retain clarity.

We assign annual fire events according to a binomial probability of a fire occurring in the present year given the last fire and difference between the years, with the base probability of fire occurring on any given year set to 1/FRI. However, we wanted to check what happens if this assignment rule were to be changed and an alternate fire regime adopted. To do this, we compare the described fire assignment with an assignment that sets the fire fraction equal to 1/FRI annually. This results in a portion of the overall area always being assigned a fire event every year bypassing the wait time of the binomial assignment, therefore, we expect more fires to be assigned according to this regime. As expected, we see that the latter fire regime worsens the overall achievable TEC (Fig. 15c) and also shows very high variance (see

Fig. 16c). Moreover, the latter regime also alters the slope of the TEC with the afforestation area, changing how much TEC can be retained at scale when this fire regime is applicable (Fig. 15d). As the perfect modeling of the actual fire regime is not possible in practice, these two regimes allow us to observe the expected ranges. We expect that the actual fire regime might be between the range spanned by these two regimes. We also observe that changing the fire fractions from 1% to 4% doesn't alter the TEC by much (Fig. 16d) when we have the binomial assignment, pointing to the fact that a more stricter fire assignment is needed to see noticeable changes with fire fractions.

## Modulating Forest Density Using Mortality as a Surrogate

The density of the forest, as determined by parameters such as diameter at breast height (dbh), stocking percent, and number of trees per given area, greatly affects growth of the individual trees and therefore the carbon sequestered [1, 2]. However, exactly how yield changes with the notion of density depends on variables such as site conditions and species composition, and there is no reliable universal relationship [3, 4]. One could potentially calibrate a growth and yield model (such as GYPSY [5], TIPSY [6], or VDYP [7]) that can simulate yields for varying density settings with data from the region of interest, however, without such calibration data, even such a simulation would provide a range of possibilities that need to be carefully narrowed using available heuristics. We intend to explore using such a growth and yield model in a future study, but for this study, we use generic mortality to modulate the density parameter.

The common way to simulate the setting with density is to obtain the yield curves for a plot with a specific density of trees. The yield curves we use in this study are a result of fitting yield equations to sample plots across Canada, and therefore can be thought of as applicable for a density that is an average of the ones seen across these sample plots. Considering this as the average density, we can then modulate the notion of density using mortality to control what percentage of trees are allowed to remain. For instance, in our region of interest near the edge of the treeline, we can expect that the trees are present with less than average density. We simulate this setting by introducing mortality of existing trees on FL in the year 2020 and then conducting afforestation with average density in the year 2025, resulting in a configuration of filling an existing sparsely populated FL.

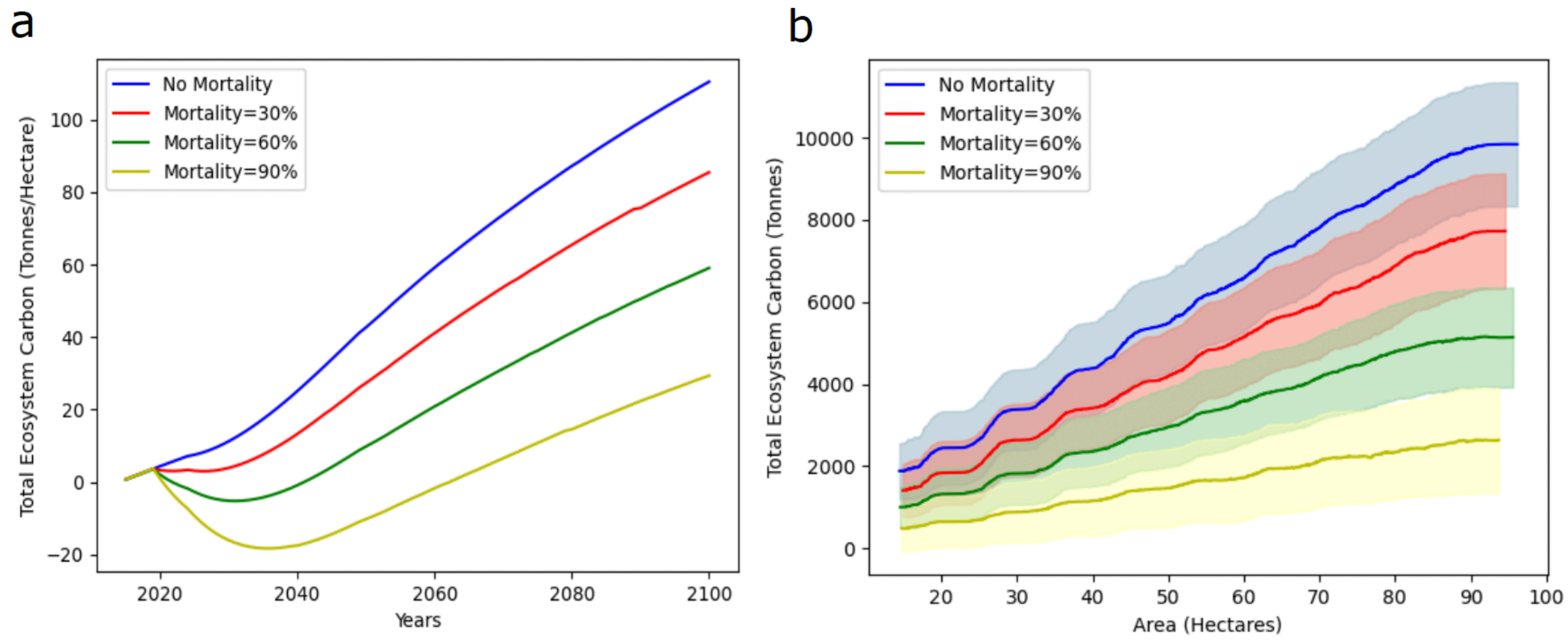

Figure 17: **TEC for varying mortality of existing trees on forested land (FL). a)** TEC over the years. The standard deviations can be found in supplementary Fig. 11. **b)** TEC at 2100 as a function of the afforestation area. A combination of existing simulations is sampled and combined to get higher afforestation area and the resulting TEC is added. The lines denote the mean and the spread shows the standard deviation. The TEC accounts for existing plus afforested trees on FL.

We see that mortality of existing trees (density reduction) changes overall TEC growth rate over the years (Fig. 17a) and across larger areas (Fig. 17b). If our assumptions about mortality and density hold, these results point to the fact that by planting with average density on existing sparse forests, the total capture capacity at 2100 falls linearly with existing density (Fig. 17a,b). This means that in a really sparse forest, the density of afforested trees need to compensate linearly for the sparsity of existing trees, in order to reach the capture capacity of a forest with average density. Nonetheless, this does not factor in how existing density can change the growth patterns of afforested trees and how mixture of species behave together, for which, the aforementioned growth and yield model simulation with location, species, and density specific calibration may be necessary. Another avenue for investigating the effect of existing and afforested density on captured carbon is by using yield curves that factor in varying density as distinct scenarios in your region of interest, as in the yield curves produced by Timberworks Inc. in NT [8]. We study these yield curves in the main article.

## Future Research

Future research could expand the analysis to include the Taiga Cordillera ecozone, improving coverage of north-western landscapes and capturing ecozone variability. A key step will be incorporating province-specific yield curves for even more accurate growth estimations, supplemented by systematic exploration of parameter spaces in tools such as GYPSY [5] and TIPSY [6] to represent stand densities and climate sensitivities. To generate a high-resolution carbon capture map as a decision-support resource, transitioning to full-scale GCBM simulations and obtaining monte-carlo carbon capture estimates for each cell in the region of interest may be necessary. Moreover, modeling historical insect disturbances and overlaying them onto current stand conditions can help refine predictions where pest outbreaks pose a significant risk to newly established forests. In parallel, coupling climate-projection data with advanced disturbance models for wildfires and insect outbreaks would illuminate how changing conditions amplify or mitigate these risks. Additionally, exploring diverse forest management scenarios, ranging from partial cutting to salvage logging and strategic harvesting, within these modeling frameworks could help identify the best pathways to balance carbon gains and ecological resilience. Studying scenarios that employ well-timed harvesting regimes to minimize stand senescence and mortality-related losses, while performing afforestation in these places, could further enhance the net climate benefits of afforestation initiatives. Integrating explicit soil data from the National Soil Database (NSDB) [9] will sharpen our grasp of site-level constraints, including those in the southern Arctic transition zone where rapid warming could accelerate or hinder afforestation. However, modeling expansions into cryosolic soils at the southern Arctic edge presents an additional hurdle, as GCBM may not currently support these substrates, and few, if any, yield curves exist for such cold, soil-limited conditions. Finally, incorporating alternative fire regime models that account for stand age classes and potential propagation to neighboring cells (e.g., Burn-P3 [10]) will provide a more nuanced representation of disturbance patterns, thereby offering improved estimates of long-term carbon storage under large-scale boreal and taiga afforestation projects.